\documentclass{article}

\usepackage{amsmath,amssymb}
\usepackage{subfigure}
\newcommand{\independent}{\rotatebox[origin=c]{90}{$\models$}}
\usepackage{graphicx} 
\usepackage{color}
\usepackage[a4paper,left=0.85in,right=0.85in,top=1in,bottom=1in]{geometry}
\usepackage{hhline}
\usepackage{multirow}

\usepackage{url}
\usepackage{breakurl}
\usepackage[breaklinks]{hyperref}

\RequirePackage{natbib}

\title{\sc{Dirichlet Process Mixtures of Order Statistics with Applications to Retail Analytics}}
\author{James Pitkin, Gordon Ross and Ioanna Manolopoulou}

\begin{document}
\maketitle

\begin{abstract}
 The rise of ``big data'' has led to the frequent need to process and store datasets containing large numbers of high dimensional observations. Due to storage restrictions, these observations might be recorded in a lossy-but-sparse manner, with information collapsed onto a few entries which are considered important. This results in informative missingness in the observed data. Our motivating application comes from retail analytics, where the behaviour of product sales is summarised by the price elasticity of each product with respect to a small number of its top competitors. The resulting data are vectors of  order statistics, due to only the top few entries being observed. Interest lies in characterising the behaviour of a product's competitors, and clustering products based on how their competition is spread across the market. We develop nonparametric Bayesian methodology for modelling vectors of order statistics that utilises a Dirichlet Process Mixture Model with an Exponentiated Weibull kernel. Our approach allows us added flexibility for the distribution of each vector, while providing parameters that characterise the decay of the leading entries. We implement our methods on a retail analytics dataset of the cross-elasticity coefficients, and our analysis reveals distinct types of behaviour across the different products of interest.
\end{abstract}

\section{\label{sec:Introduction}Introduction}
The field of retail analytics is concerned with understanding the purchasing behaviour of consumers, for purposes such as sales prediction, inventory management and coupon personalisation \citep{silver2013concurrent, rudin2013learning, gunawardana2009survey, Huh_Rusmevichientong_2009, bajari2015machine, ferreira2015analytics}. Retail analytics is a particularly challenging area for machine learning, since vast amounts of data are collected at various operational levels, ranging from individual customer transaction data, to aggregated sales data across whole sectors. This means that companies are interested in developing efficient summaries of their data, to mitigate storage and computational costs \citep{akcay2013statistical, Intel_retail}. 

A particularly important example involves the price-elasticity coefficients generated by sales prediction models. Given a set of products which are offered to consumers, the sales of each product typically depend on both its own price as well as the price of each of its competitors. The cross-elasticity of two products is a measure of the dependence that their prices have on their respective sales. In theory, companies would store this information as a matrix which contains the cross-elasticities for each pair of products. However, in practice, computing price elasticities can be computationally prohibitive when there are a large number of products,  so companies often instead make use of highly tailored black-box sparse regression sales models \citep{liu2013sales, beheshti2015survey} and only measure the cross elasticity for a small number of each product's competitors, with the remaining entries of the matrix treated as missing or negligible.
 
Although the resulting output matrix of coefficients can be successful at providing accurate predictions of future sales, the inherent informative missingness implies that a global interpretation and understanding of the behaviour of the market may not be directly available. Here we are particularly interested in clustering groups of similar products together according to the distribution of their competition in the market, but also in identifying products for which potentially important competitors may have been missed out of the observed cross-elasticity matrix.

Formally, the form of the cross-elasticity data at hand is such that, for each product $i$, we observe a decreasing set of entries (i.e.~observed order statistics) of a larger vector, that have been censored for sparsity purposes to only the top few entries. 
Mathematically speaking, the data are in the form:
$$\textbf{\mbox{X}} = \{  \boldsymbol{x}_{i,1:n} : x_{i,n-l_i+1} \leq x_{i,n-l_i+2}\leq\ldots\leq x_{i,n}, \text{with } x_{i,j}  \text{ censored to 0 for $ 1\leq j \leq n-l_i$} \},$$
where $\boldsymbol x\rm_{i}  $ is the cross elasticity vector of dimension $n$ for product $i$, which has $l_i$ uncensored ordered entries, with the remaining being censored. Heterogeneity among products stems both from the rate in which competition decays, as well as the actual number of uncensored cross-elasticities. Existing price sensitivity analyses focus on reducing these vectors to summary statistics of elasticity coefficients from fitted demand models \citep{andreyeva2010impact, oliveira2007consumer}. To our knowledge, there is currently no methodology to cluster these entire vectors with minimal information loss, which would allow the flexibility in handling the varying lengths of cross-elasticity vectors as well as incorporating the censoring mechanism to provide information about the censored entries. 

In this paper we develop non-parametric Bayesian models by interpreting our observed data as realisations of \it variable length order statistics sequences\rm. We will show that this succinctly handles the partial censoring and allows for computationally straight-forward inference on the unobserved entries of the cross-elasticity matrix. Our approach uses tools from survival analysis to address inherent censoring mechanisms, together with non-parametric Bayesian Dirichlet Process mixture models that allow products to be clustered into distinct groups for the purpose of analysis. Using the Exponentiated Weibull distribution as a mixture kernel \citep{mudholkar1993exponentiated}, we are able to account for both light and heavy tail behaviour apparent in the data. As we will discuss later, the Exponentiated Weibull distribution has several unique properties which make it ideal for modelling order statistics. We develop efficient sampling mechanisms by adapting  algorithm 8 of \cite{neal2000markov} and provide interpretation and visualisation tools for summarising and presenting the output. Our approach fully characterises sales sensitivities by incorporating all the information from the entire cross-elasticity vector, offering two distinct benefits. Firstly, by interpreting these elasticity vectors as order statistic sequences we can directly cluster products in terms of their entire cross elasticity vectors and  conveniently handle their varying length nature. Secondly, it provides a framework for predicting censored entries which can shed light on potentially important competitors which have been censored. 

Although we focus on the retail analytics application, our methodology is general and is relevant in any situation with informative missingness where only the top few order statistics of each observation vector are observed. This includes applications such as sports analytics, where only the top few performances, athletes or teams are observed \citep{Malcata2014}, or the stylometry analysis of literary texts which often focuses on analysing the frequencies of the top few most common words \citep{narayanan2012feasibility}.

The rest of the paper is organised as follows: Section \ref{sec:Motivation} outlines our motivation for clustering elasticity coefficients in the retail analytics setting. Section \ref{sec:OrderCont} covers the properties of uniform order statistics relevant to our model and  reviews the relevant literature, in particular drawing parallels to survival and reliability analysis. Section \ref{sec:ExpWeib} provides a background of the pertinent characteristics of the Exponentiated Weibull distribution and its relevance as a kernel to \it variable length order statistics sequences\rm. Section \ref{sec:NPVLOSS}  covers the theory of Dirichlet process mixture models and further provides the nonparametric mixture model of \it variable length order statistics sequences \rm along with prior distributions specification. We outline the algorithm used in posterior inference in Section \ref{sec:posterior_inference}. Section \ref{sec:DataExamples} illustrates our methods on a simulated and a real retail analytics dataset of cross-elasticity vectors. Dunnhumby Ltd, a customer science company, allowed us  access to  the anonymised cross-elasticity coefficient output of a set of products derived from the loyalty card transactions of leading UK supermarket retailer. Section \ref{sec:Conclusion} provides a summary of our methods with potential extensions and applications of the work. 
  
\section{\label{sec:Motivation}Motivation}


It is common in retail analytics to characterise products based on how sensitive their sales are to the prices of their competitors and how customers interact with their product range. Analytics teams are constantly striving to develop models and inference methods that provide insight into understanding how price fluctuations that propagate throughout stores will impact the sales of products whose prices have not changed \citep{persson1995modeling, ferreira2015analytics}. Clustering products on the basis of their price sensitivity profile can provide a segmentation of a retailer's product range. This ultimately aids store planners in deciding on the value of a given pricing or display combination, as it provides information on how a product's sales are likely to react to the deviations of prices of other products. For instance, a poor display combination could be one that consists entirely of products characterised by their sales being primarily  driven by the prices of its competition. This would lead to margin cannibalisation - where profit made on one product is offset by the loss of profit of another product. The information of product clusterings would allow for better  pricing and display optimisation.

One approach to such a sales sensitivity analysis is to cluster products in terms of their direct and cross-elasticity coefficients. 
Existing work on analysing sales sensitivities of retail goods, and using these to partition product ranges, has focused on defining summary statistics which capture many of the important aspects of price sensitivity profiles \citep{andreyeva2010impact}. For example, \cite{oliveira2007consumer} investigated the heterogeneity in direct (but not cross) elasticities across products and across consumer groups.
Similar work investigated variations in category-level summaries of cross-elasticities coefficients and the impact of sales sensitivity across store, demographic and product category levels  \citep{hoch1995determinants,guerrero2017price}. However, much of the important information of price sensitivity profiles lies in the entire cross-elasticity vectors and cannot be captured in summary statistics. 

 The dataset we have access to through dunnhumby ltd comprises cross-elasticity vectors for a set of products from a leading UK supermarket chain. Although the precise mechanics of how these estimates are obtained are highly engineered within their proprietary model, the general form of the model is given by refined versions of the Working-Leser regression \citep{working1943statistical,leser1963forms}:
\begin{equation} \label{Price_strat_desc}
  \log \left( S_{i,t} \right)  =  -\varphi_{i} \log \left(Q_{i,t} \right)+\sum_{j=1}^{n_i}\varphi_{i}\eta_{ij}\log\left(P_{i,j,t}\right)+ f\left(Q_{i,1:T},P_{i,1:n_i,1:T}\right)+\epsilon_{i,t},
\end{equation}
where, for each product $i$ and discrete time $t$ (in days), $S_{it}$  denotes its sales, $Q_{it}$ its price, $P_{ijt} $ the price of its  $j^{th}$ competitor product, $\varphi_{i} $  its direct elasticity and $\eta_{ij} $ product $j$'s relative cross elasticity with product $i$ (as a multiple of the direct elasticity). The term $\epsilon_{i,t}$ represents time-dependent error. We use the notation $1:n$ to denote the set $1,\ldots,n$. The function $f(\cdot)$  involves data aggregation and seasonality patterns relevant to retail sales, as well as additional information on display combinations and promotions; our focus here is not on this function, but rather on the post-processing of the output of the regression model. 
Here $n_i$ is the number of competitor products of product $i$, which are pre-selected using expert knowledge encoded in an algorithm, to avoid using the entire set of products which is computationally prohibitive due to the complexity of $f(\cdot)$. 
For the purposes of this study and to ease notation in later sections, we assume that competitor products are labelled such that product $i$'s cross-elasticity coefficients $\eta_{ij}$ are increasing in magnitude and that all products have the same potential number of competitors, i.e.~$n_i=n\;i=1,\ldots,N$.  The cross-elasticity coefficients are estimated using shrinkage methods for sparsity reasons, so that only $l_i$ $\eta$'s are non-zero, with the remaining exactly equal to 0. Table \ref{table:Elasticity_examples} provides two toy examples of \textit{variable length order statistic sequences} in the context of cross-elasticities.

\begin{center}
\begin{table}
\centering
\caption{\label{table:Elasticity_examples}  Ordered elasticity output $\varphi$ and $\eta$ for two fictional products, \textit{Bobby's puffs} and \textit{Lucan's Salted crisps}. For each product we have columns of order elasticity coefficients $\varphi_{i}$, $\varphi_{i}\eta_{ij}$ along with the respective sequences of $\eta_{ij}$, which demonstrates the decreasing nature of data from model (\ref{Price_strat_desc}). The number of potential cross competitors is set to $n_{i} =6$, although the number of terms censored to 0 differs. Importantly, the set of competitors can differ for each of the products and in instances where there is a shared competitor  (as with \textit{Supermarket puffs} in this case), the value of $\varphi_{i}\eta_{ij}$, as well as its position in the ordering, need not be consistent across products.}  
\begin{tabular}{|c|c|c|c||||c|c|c|}
  \hline
  \multirow{2}{*}{} &
      \multicolumn{3}{c||||}{\bf{Bobby's Cheesy puffs} } &
      \multicolumn{3}{c|}{\bf{Lucan's Salted crisps}} \\	\cline{2-7}
   & Relevant competitors & $\varphi_{1},\varphi_{1}\eta_{1j}$ & $\eta_{1j}$ & $\varphi_{2},\varphi_{2}\eta_{2j}$ & $\eta_{2j}$ & Relevant competitors  \\ 
  \hhline{|=|=|=|=||||=|=|=|} \hhline{|=|=|=|=||||=|=|=|}
  $\varphi_{i}$ & \text{Bobby's puffs} & -1.41 & & -1.86 & & \text{Lucan's Salted crisps} \\ \hline  
  $\varphi_{i}\eta_{i6}$ & \text{Supermarket puffs } & -1.12 & 0.79 & -0.8 & 0.43 & \text{Sussex's Chives crisps} \\ \hline
  $\varphi_{i}\eta_{i5}$ & \text{Harry's puffs} & -1.10  & 0.78 & -0.44 & 0.23 & \text{Chef's Paprika crisps} \\ \hline
  $\varphi_{i}\eta_{i4}$ & \text{Supermarket Nuts} & -0.80 & 0.57 & -0.10 & 0.05& \text{Supermarket puffs }\\ \hline
  $\varphi_{i}\eta_{i3}$ & \text{Bobby's Tortillas } & -0.48 & 0.34 &  -0.04 & 0.02 & \text{Lucan's nuts}\\ \hline
  $\varphi_{i}\eta_{i2}$ & \text{Tommy's chips } & -0.35 & 0.25 & 0 & 0 &  \text{Harry's Popcorn}\\ \hline
  $\varphi_{i}\eta_{i1}$ & \text{Tommy's puffs } & -0.05 & 0.04 & 0 & 0 & \text{Chef's BBQ crisps} \\ \hline
\end{tabular}
\end{table}
\end{center}

Fitting this model to data provides us with a vector of cross-elasticities for each product, where some of the entries may be zero due to sparsity. Our goal is to then cluster products according to these cross-elasticity vectors. Although, in theory, one can perform clustering alongside the regression, this is computationally prohibitive in the current context because of the highly tailored model fitting involved, so we treat the regression fitting as `black-box' and work with the cross-elasticity vectors directly. To address the fact that, due to computational limitations,  competitor products are pre-selected using expert knowledge  and are subject to error, we treat the zero entries as missing minor competitors (with smaller cross-elasticity coefficients than the observed ones). This results in a clustering framework whereby observation vectors have different numbers of non-missing entries.

Since cross-elasticity vectors arise as the outcome of penalised regression, it is natural to assume that coefficients are shrunk to zero as the result of a penalisation threshold. For example, in the simplest case of best-subset selection with an orthonormal design matrix, non-zero coefficients are exactly equal to the top order statistics of the corresponding ordinary least squares estimates. With this in mind, we treat the observed non-zero cross-elasticity coefficients as the top order statistics of an underlying vector of length $n$.  The $l$ observed coefficients of the leading product competitors are thus modelled as the top $l$ order statistics of a set of $n$ independent and identically distributed observations from an unknown underlying distribution. To account for the fact that a different number $l$ of entries may be observed in each vector, we assume that $l$ also follows a probability distribution, independently of the actual entries.

\section{\label{sec:OrderCont}Order statistics of continuous distributions}
The order statistics of a random sample are the \it reordered \rm observations in terms of increasing size. More concretely, given a continuous distrbution variable $X$ and observations $ x_{1:n} \overset{i.i.d.}{\sim} X $, the order statistics $x_{\left( 1 \right)},\ldots,x_{\left( n \right)} $ are given by:  
\begin{equation} \label{Order_statistics_def}
  x_{\left( 1\right)} < x_{\left( 2 \right)}<\ldots<x_{\left( n \right)}.
\end{equation}  
The $j^{th} $ order statistic of (\ref{Order_statistics_def}) is denoted as $x_{\left(j \right)}$ and thus,  $x_{\left(1 \right)}$ and $x_{\left(n \right)}$ are the smallest and largest observations respectively.  Given a density function $f\left(x \right)$ of a continuous random variable $X$, the density of the $j^{th}$ order statistic $x_{\left(j\right)} $, denoted by  $f_{\left(j \right)} \left(x \right)$ is given by \citep{arnold1992first}:
\begin{equation} \label{Order_density}
  f_{\left(j \right)}\left( x \right) = nf \left(x \right){n-1 \choose j-1}F\left(x \right)^{j-1}(1-F\left(x \right))^{n-j}.
\end{equation}  

One of our key modelling assumptions is that a partially observed cross-elasticity vector of length $n$ with $l$ non-zero entries in fact corresponds to the top $l$ order statistics of a random sample of size $n$. We term each of these vectors of the top $l$ order statistics as \it variable length order statistics sequences\rm, and denote them as $\boldsymbol{x}  = \left(x_{\left(n \right)},\ldots,  x_{\left(n-\left(l-1 \right) \right)}\right)$. We also denote the $j^{th}$ order statistic of sequence $\boldsymbol{x}$ by  $x_{\left(j \right)}$. The density of $ \boldsymbol{x}\mid l$ is denoted as $f_{\left(n \right):\left(n-l+1 \right)}$ and  given by:
 \begin{equation} \label{top_Kth_joint}
\begin{split}
  f_{\left(n \right):\left(n-l+1 \right)} \left(\boldsymbol{x} \mid l \right)  =&   f_{\left(n \right):\left(n-l+1 \right)} \left(x_{\left(n \right)},\ldots,x_{\left(n-l+1 \right)} \mid l  \right)  \\
   = & \frac{n!}{\left(n-l \right)!} F\left(x_{\left(n- \left(l-1 \right) \right)} \right)^{n-l} \prod_{j=1}^lf \left(x_{\left(n+j-l \right)} \right).  \\
\end{split}
\end{equation}
By the independence of $x_{\left(n-j \right)} \mid  x_{\left(n-j+1 \right)}  \independent  x_{\left(n \right)}, x_{\left(n-1 \right)},\ldots, x_{\left(n-j+2 \right)}  $ and by (\ref{top_Kth_joint}), the density of the conditional distribution of $x_{\left(n-j \right)} \mid  x_{\left(n-j+1 \right)},l $ for $j<l$ (denoted as $f_{ \left(n-j \right)\mid  \left(n-j+1 \right)} $) is given by:
\begin{equation} \label{conditional_order_stats}
f_{ \left(n-j \right)\mid  \left(n-j+1 \right)} \left(x_{ \left(n-j \right)} \mid  x_{\left(n-j+1 \right)},l \right) = \left(n-j \right) f \left(x_{\left(n-j \right)} \right)\frac{ F \left(x_{\left(n-j \right)}\right)^{n- \left(j+1 \right)}}{ F\left(x_{\left( n-j+1 \right)}\right)^{n-j}}
\end{equation}
and thus the  density of the joint sample $\boldsymbol{x} \mid l$ can also be expressed in hierarchical format: \begin{equation} \label{hierarchical_order_stats}
\begin{split}
f_{\left(n \right):\left(n-l+1 \right)}\left(\boldsymbol{x} \mid l \right)= f\left(x_{(n)}\right)\prod_{j=1}^{l-1} f_{(n-j)\mid (n-j+1)}\left(x_{(n-j)}\mid x_{(n-j+1)}, l\right)
\end{split}
\end{equation}
Finally,  the joint distribution of $\boldsymbol{x}$ can be combined with the $n - l$ zero entries of $\boldsymbol{x}$ through
\begin{equation}
f\left( \boldsymbol{x} \right) = p(l) \times f_{\left(n \right):\left(n-l+1 \right)}\left(\boldsymbol{x} \mid l \right),
\end{equation}
where is $p(l)$ is the probability mass function over the length of the sequence. Here we assume that $l$ and the magnitude of the non-zero entries of $\boldsymbol{x}$ are independent. 

Much work has been done in the study of the theoretical properties of order statistics \citep{beutner2009order}  and has been applied to areas such as modelling software reliability  \citep{wilson2007nonparametric}, reliability of propulsion systems of aircraft \citep{warr2014bayesian} and recommender systems \citep{caron2012bayesian}. 
A particularly relevant field of order statistics which bears resemblance to our problem set-up lies in the field of reliability analysis, known as $k$-out-of-$n$ systems.  A $k$-out-of-$n$ system models the failure of $k$ out of $n$ components within a finite time horizon. The set of $k$ ordered values of the time until failure (censored or not) can then be modelled as the observed order statistics of a base distribution. Much of the relevant non-parametric work  has focused on flexibly learning the underlying base distributions \citep{wilson2007nonparametric,barghout1998non} and building hierarchical versions of these models \citep{ghosh2007nonparametric}. In the $k$-out-of-$n$ framework, a standard assumption is that each sequence/system produces the same marginal order statistic, whereas we would like to allow for additional flexibility. 

In the current context, we observe the top few order statistics of the cross-elasticity vector, with the remaining entries treated as missing. This type of data is akin to the format of models in survival analysis, where the probability of survival decreases over time and may be right-censored. One aspect important to the success of Bayesian non-parametric models in survival analysis is the choice of kernel, as it impacts whether the relevant statistics and survival functions are recoverable. As a consequence, much attention is paid to the choice of kernel. Notably, a hierarchical structure in the base measure was introduced by  \cite{de2004anova}, whereas \cite{hanson2006modeling} and  \cite{kottas2006nonparametric} used Gamma and Weibull kernels within a Dirichlet process mixture model framework respectively.  
The Exponentiated Weibull distribution was shown to be the first distribution that could model non-monotone hazards \citep{mudholkar1993exponentiated}, which in our context correspond to order statistics terms whose modes exist but are not necessarily light-tailed.

\section{\label{sec:ExpWeib}Exponentiated Weibull distribution}
Following the formulation of our observations as order statistics of random samples, the choice of the underlying distribution of $X$ will determine the behaviour of the corresponding order statistics. Here we are interested in a distribution which can allow for a range of light and heavy tail behaviour and provide interpretable analytical expressions for the distribution of its order statistics. We thus assume that these random samples are distributed according to the Exponentiated Weibull distribution. A random variable $X$ is distributed according to the Exponentiated Weibull (EW) distribution, denoted as $ X\sim EW\left(\alpha, \beta, \lambda\right)$, if its probability density and distribution function are given by
\begin{equation} \label{EW_density}
  f\left(x\right) = \alpha \beta \lambda^{\beta} x^{\beta-1} \left(1-e^{-\left(\lambda x\right)^{\beta}}\right)^{\alpha-1} e^{-\left(\lambda x\right)^{\beta}} \\	
\end{equation}    
and 
\begin{equation} \label{EW_cdf}
  F\left(x\right) =  \left(1-e^{-\left(\lambda x\right)^{\beta}}\right)^{\alpha}\\	
\end{equation}      
respectively, where $  x  >0, \lambda >0, \beta >0, \alpha >0$.

\begin{figure}[h]
  \centering
    {
      \includegraphics[width=0.5\textwidth]{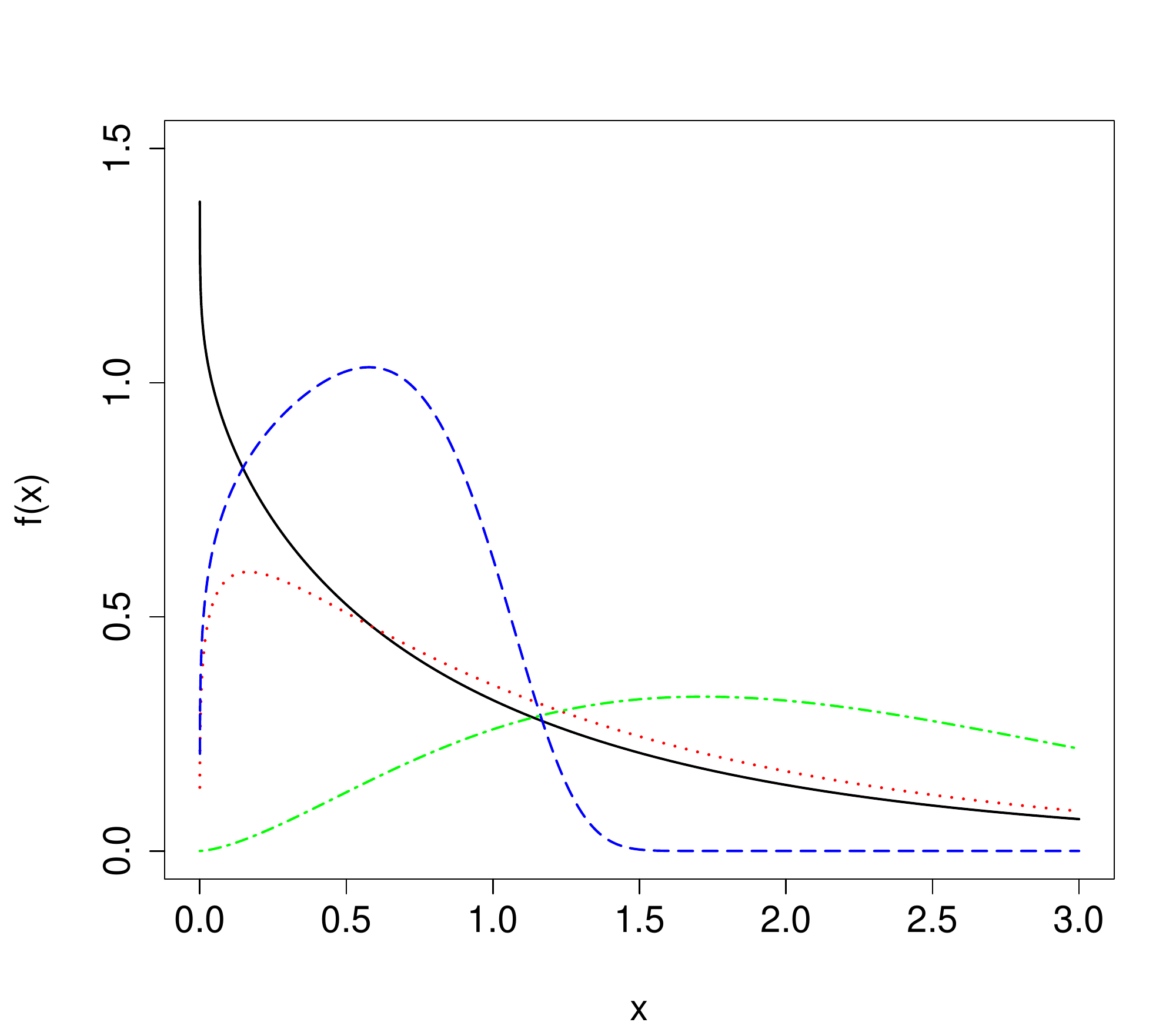}
    }
  \caption{EW density for $\left(\alpha, \beta, \lambda\right)$ = (1.2, 0.8, 1.0) [black solid], (1.55, 0.8, 1.0) [blue dashed], (0.24, 5.0, 1.0) [red dotted] and (1.8, 1.4, 0.5) [green dashed-dotted lines] respectively.}\label{plot:EWdensities}
\end{figure}

The Exponentiated Weibull is an extension to the standard Weibull distribution through the inclusion of the additional parameter $\alpha$, which allows the distribution to have a wide range of tail behaviours. Similarly to the Weibull distribution,  $\lambda$ is a scale parameter whereas $ \beta $ controls the tail behaviour of the distribution; distributions are heavy tailed for $ \beta <1$ and light-tailed otherwise. Furthermore, decreasing $\beta$ monotonically increases the mean and variance, kurtosis and skew of the EW distribution. The impact of $ \alpha$ depends on both the value $\alpha\beta $ and whether $ \alpha <1$;  increasing $\alpha$  increases symmetry around the mean and mode. 
These different modal, asymptotic and tail behaviours  \citep{nassar2003exponentiated} are summarised in Table \ref{table:EWparams}. 
Figure \ref{plot:EWdensities} demonstrates various density plots for differing combinations of $\left(\alpha, \beta, \lambda\right)$, various asymptotic, modal and tail behaviours are observed.   

\begin{center}
\begin{table}
\centering
\caption{\label{table:EWparams}  EW density behaviours for various combinations of $\left(\alpha, \beta, \lambda\right)$  }  
    \begin{tabular}{| l | l | l | l |}
      \hline
      Ranges of $\alpha, \beta $ & $ x \rightarrow 0$ &   Mode   & Order statistic marginal tails\\ \hline
      $\alpha  >1,  \beta >1,  \alpha\beta  > 1$  & $f\left(x\right) \rightarrow 0$ &  $\approx \frac{1}{\lambda}  \left[\frac{2\left(\alpha\beta-1\right)}{\beta\left(\alpha+1\right)}\right]^{1/ \beta} $  &  Light\\ 
      $\alpha  > 1,  \beta < 1, \alpha\beta >1$   & $f\left(x\right) \rightarrow 0$ & $ \approx \frac{1}{\lambda}  \left[\frac{2\left(\alpha\beta-1\right)}{\beta\left(\alpha+1\right)}\right]^{1/ \beta} $ & Heavy\\
      $\alpha  > 1,  \beta < 1, \alpha\beta <1$   & $f\left(x\right) \rightarrow \infty$ &  none  & Heavy\\
      $\alpha  < 1,  \beta > 1, \alpha\beta <1$   & $f\left(x\right) \rightarrow \infty$ &  none  & Light\\      
      $\alpha  < 1,  \beta > 1, \alpha\beta = 1$   & $f\left(x\right) \rightarrow \lambda$ & 0 & Light\\
      
      \hline
    \end{tabular}
\end{table}
\end{center}

\subsection{EW distribution application to order statistics}\label{subsec:EW_appto_OrdStat}	
There are some key properties of the EW distribution that lead to useful applications to order statistics and \it variable length order statistics sequences\rm. The joint density of (\ref{top_Kth_joint})  under the EW distribution for fixed order sequences of lengths $l$ is given by:
\begin{equation} \label{EW_joint_VLOSS_density}
  f_{\left(n\right):\left(n-l+1\right)} \left(\boldsymbol{x} \mid l \right)  = \frac{n!}{\left(n-l\right)!} \left(1-e^{-\left(\lambda x_{\left(n-\left(l-1\right)\right)}\right)^{\beta}}\right)^{\alpha\left(n-l\right)}  \prod_{j=1}^lf\left(x_{\left(n+j-l\right)}\right)
\end{equation}
where $f$ is the EW density function of (\ref{EW_density}). 

\begin{figure}[h]
  \centering
     \includegraphics[width=0.45\textwidth]{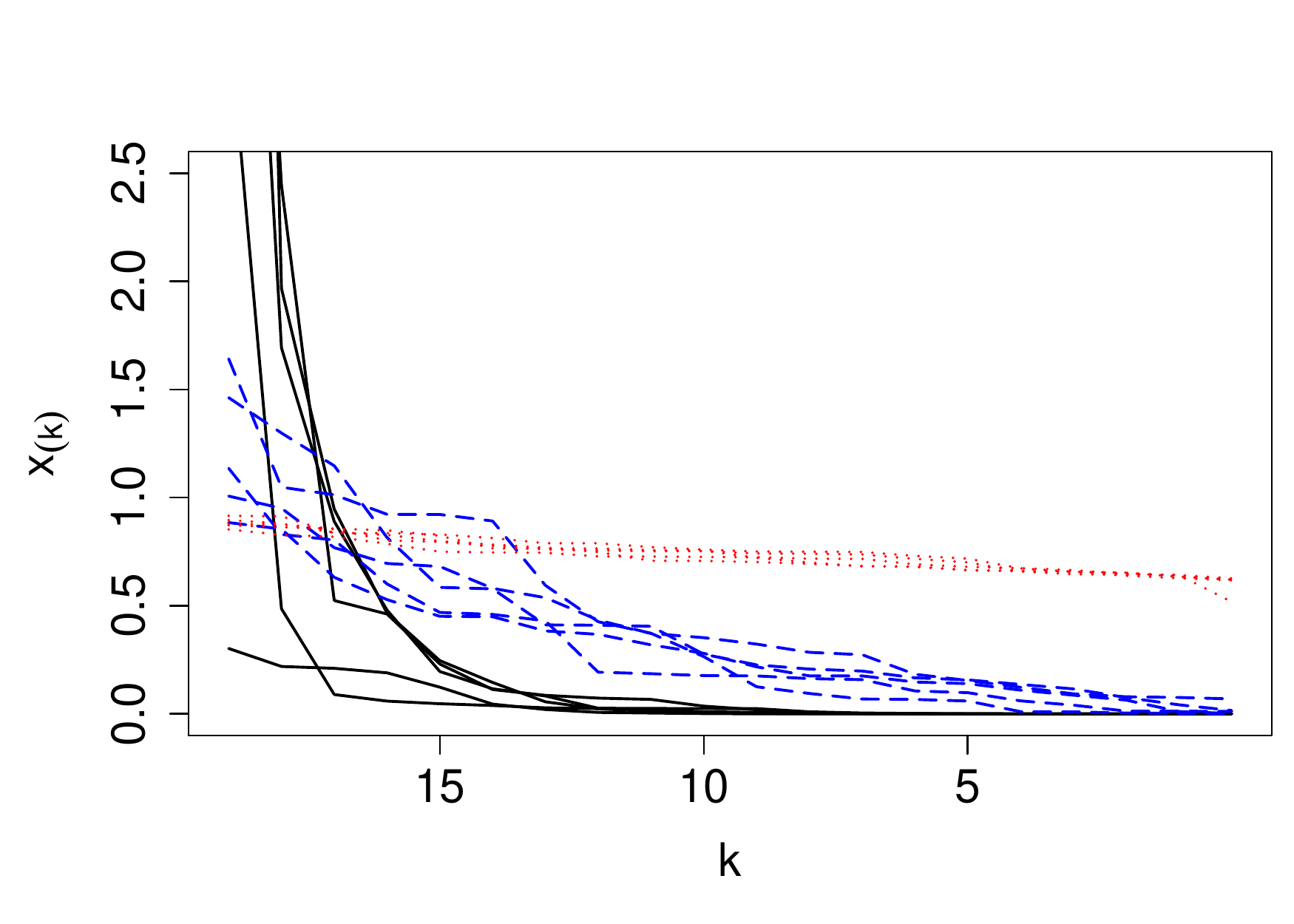}
     \includegraphics[width=0.45\textwidth]{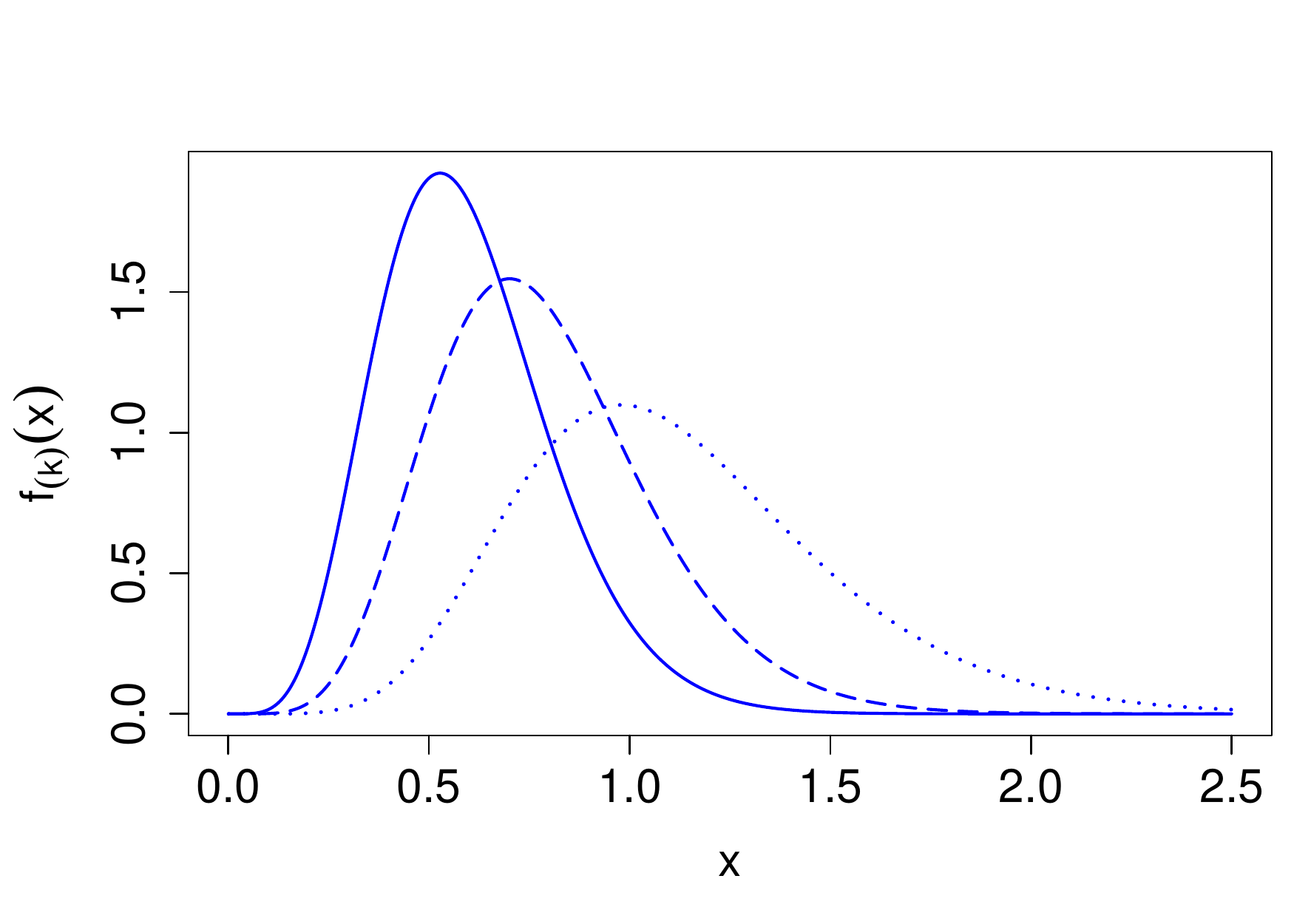}
    \caption{Left panel: realizations of order statistics sequences with $EW\left(\alpha, \beta, \lambda \right)$ kernel for combinations $\left(\alpha, \beta, \lambda \right)$ = (0.2, 0.6, 0.7) [black solid], (0.5, 1.5, 1.5) [blue dashed], (4, 5, 1.5) [red dotted] respectively. Right panel: Density plots of $f_{\left(k \right)} \left(x \right)$ with EW(0.5, 1.5, 1.5) kernel for orders $ k$=10 [dotted], 9 [dashed] and 8 [solid] respectively.}\label{plot:OrderStatSeq_OrderMarginalDensity}
\end{figure}

The EW distribution handles censoring naturally, since the censored, joint and conditional densities under the EW distribution belong to the same family. For example, if $ x_i \overset{iid}{\sim} EW\left(\alpha,\beta,\lambda\right) $ for $i=1,2,\ldots,n$, then  $x_{\left(j\right)} \sim EW\left(j\alpha, \beta, \lambda \right) $.   Similarly, the conditional distributions $x_{\left(n-j\right)}  \mid  x_{\left(n-j+1\right)}  \sim EW_{ x_{\left(n-j\right)}  <x_{\left(n-j+1\right)}}\left(\left(n-j\right)\alpha, \beta, \lambda\right), 1 \leq j \leq n-1 $ are also readily available. This means that the properties and interpretability of the EW distribution transparently carry over to its order statistics. Finally, the EW can account for both light and heavy tails, allowing us to capture different types of decay behaviours of the elasticity vectors. Figure \ref{plot:OrderStatSeq_OrderMarginalDensity} provides some examples of order statistics sequences, which demonstrate various decay behaviours and tail behaviours that can be produced under the EW kernel.

\section{\label{sec:NPVLOSS}Nonparametric mixture model of variable length order statistics sequences}
As outlined in Section \ref{sec:Motivation}, our ultimate goal is to cluster variable length order statistics sequences, here arising through the behaviour of the cross-elasticity vectors of different products. To this end, we use the EW distribution as a representation of cross-elasticity decay behaviour. However, in order to account for different behaviour across products, we additionally cluster products that potentially correspond to the same EW distribution. We opt for a mixture modelling framework to allow fully model-based uncertainty to propagate through the clustering inference. To account for an unknown number of underlying components, we use a Bayesian non-parametric mixture modelling formulation, guiding the number of clusters through a prior on the base distribution of a component. We thus model the entire set of cross-elasticity vectors non-parametrically as a Dirichlet Process Mixture Model \citep{antoniak1974mixtures}. 

A Dirichlet process (DP) is a distribution over random probability measures \citep{ferguson1973bayesian}, parameterised by base distribution $G_0$ over a measurable space $\Theta $, and concentration parameter $ \nu$, denoted as $G\sim  \text{DP}\left(\nu G_0\right) $. Realisations from DP$\left(\nu G_0\right)$ are \it centred \rm around the base distribution $G_0$, that is, for any measurable set $ A \subset \Theta$, $\mathbb{E}\left[G(A)\right] = G_0\left(A\right)$. The concentration parameter $\nu $ controls the degree to which realisations from DP$\left(\nu G_0\right)$ are \it close \rm to $G_0$.  \cite{sethuraman1994constructive}  established a convenient formulation of a DP known as a stick-breaking construction, which expresses a distribution $G \sim  \text{DP}\left(\nu G_0\right) $ as 
\begin{equation} \label{stick_breaking}
  \begin{split}
    G & = \sum_{i=1}^{\infty} \pi_{i} \delta_{\theta_i}, \\
    \beta_{i}  & \overset{i.i.d.}{\sim}   Beta\left(1, \nu\right), \\
    \pi_{i} = & \beta_{i} \prod_{j=1}^{i-1}\left(1-\beta_{j}\right),
  \end{split}
\end{equation}
where $ \delta_{x}$ is the Dirac measure of mass centred at $x$ and $ \theta_{i} \overset{i.i.d.}{\sim} G_0$. A Dirichlet process mixture model (DPMM) was proposed by \cite{antoniak1974mixtures} as a mixture model with a DP prior over the random mixing distribution. A DPMM can be expressed hierarchically:
\begin{equation} \label{DPMM_hierarchical}
  \begin{split}
    x_{i} \mid \theta_i & \overset{ind.}{\sim} \boldsymbol{\pi}\left(x_{i} \mid \theta_{i}\right),\ i=1,\ldots,N \\
    \theta_i \mid G &   \overset{i.i.d.}{\sim}  G  \\
    G \mid \nu, \omega  & \sim   \text{DP}\left(\nu G_0\right);\ G_0 = G_0\left(. \mid \omega\right) \\
    \nu, \omega & \sim   F_{1} \times F_{2}
  \end{split}
\end{equation}
where $ \boldsymbol \pi \rm$ is the response distribution of $x$ and $F_{1}, F_{2} $ are independent priors of parameters $ \nu, \omega $ respectively. The unique values of a vector $\theta$ are referred to as $\theta^*$ and use $\theta^*_{C_i}$ to denote $\theta_i$ given an allocation of observation $i$ into cluster $C_i$. 
DPMM's can be used to estimate a density $p\left(x\right)$ by a countably infinite mixture of kernels functions by placing a DP prior over the mixing distribution. Thus given a density $p\left(x\right)$ to estimate, and the family of density $f\left(x \mid \theta\right)$ parametrised by $ \theta$, then:
\begin{equation} \label{density_est}
  \begin{split}
    p\left(x\right)   & = \int f\left(x \mid \theta\right) dG\left(\theta\right) \\
    & = \sum_{i}^{\infty} \pi_{i}f\left(x \mid \theta^*_{i}\right)
  \end{split}
\end{equation}
since  $G $ omits to the representation $G = \sum_{i=1}^{\infty} \pi_{i} \delta_{\theta_i}$ with $G \sim$ DP$\left(\nu G_0\right) $.

\subsection{The model}\label{subsec:NPVLOSSmodel}
We now propose a DPMM of \it variable length order statistics sequences \rm  on mixtures of distributions satisfying (\ref{EW_joint_VLOSS_density}). Placing a DP$\left(\nu G_0 \right)$ on the distributions of (\ref{EW_joint_VLOSS_density}) is an attractive approach to handling the complex multi-modalities, decay rates and variable lengths that order statistics sequences can exhibit as discussed in Section \ref{sec:ExpWeib}. Thus, the DPMM of \it variable length order statistics sequences \rm expressed in hierarchical format of (\ref{hierarchical_order_stats}) by:
\begin{equation} \label{DPMM_hierarchical_order_sequences_EW}
  \begin{split} 
 \nu  & \sim  Gamma\left( \tau_1, \tau_2\right), \\
 G \mid \nu & \sim \text{DP}\left(\nu G_0\right), \\
 \left(\alpha_i,\beta_i,\lambda_i,w_i\right) \mid G & \sim G, \\
 l_i & \sim 1+Binomial\left( n-1, w_i\right),\\
x_{i,j} &\sim EW\left(\alpha_i,\beta_i,\lambda_i\right),\;j=1,\ldots,n,     
  \end{split}
\end{equation}
where $ i=1,2,\ldots,N$ are the number of observations and for each observation vector $i$, with all but the top $l_i$ entries being censored. The final line of (\ref{DPMM_hierarchical_order_sequences_EW}) can also be expressed through the iterative formulation:
\begin{equation}
\begin{split}
 {x}_{i,\left(n-j\right)}  \mid  {x}_{i,\left(n-j+1\right)} & \sim EW_{ {x}_{i,\left(n-j\right)}  <{x}_{i,\left(n-j+1\right)}}\left(\left(n-j\right)\alpha_i, \beta_i, \lambda_i\right), 1 \leq j \leq l_i-1  \\
    {x}_{i,\left(n\right)} & \sim EW\left(n\alpha_i, \beta_i, \lambda_i\right).  
\end{split}
\end{equation}
which follows from equation (\ref{hierarchical_order_stats}). We treat the lengths $l$ and observations ${x}_{i,\left(j \right)}$ of $\boldsymbol{x}$ as independent to allow detection of competitor omissions and to ease computation. Since cross-elasticity coefficients are identically distributed a priori, each individual coefficient has the same probability of being censored, leading to a Binomial prior on $l_i$; to avoid the degenerate case of empty cross-elasticity vectors, we force one of the Bernoulli trials to be 1. 

The base distribution $ G_0$ is a key aspect of the DP$\left(\nu G_0\right)$ as it specifies the prior over $\left(\alpha,\beta,\lambda, \omega\right)$ atoms which  defines the cluster structure of the model; here we specify $G_0$ as
\begin{eqnarray}\label{DPMM_Base}
G_0\left(\alpha,\beta,\lambda,w\right) &=&  Gamma\left(\alpha \mid \alpha^1,\alpha^2\right)  \times Gamma\left(\beta \mid \beta^1,\beta^2\right)  \times \nonumber\\
&&\phantom{ Gamma\left(\beta \mid \beta^1,\beta^2\right) }\times Gamma\left(\lambda \mid \lambda^1,\lambda^2\right) \times   Beta\left(w \mid a, b\right). 
\end{eqnarray}
The hyperparameters $\left(a, b, \alpha^1, \alpha^2, \beta^1, \beta^2, \lambda^1, \lambda^2\right) $ are treated as fixed, chosen depending on the modelling context and reflecting prior expertise.
The prior for $\nu$ is assumed to be $Gamma\left(\tau_1, \tau_2 \right)$, allowing the relation $\mathbb{E}\left[N^{*} \mid \nu\right] = \nu log\left(\frac{\nu+N}{\nu}\right)$  \citep{escobar1995bayesian} (where $N^{*}$ is the number of occupied clusters) to inform our  prior expectation of the number of clusters.

\section{Posterior inference}\label{sec:posterior_inference}
We now present an efficient Markov Chain Monte Carlo (MCMC) procedure for obtaining samples from the posterior of $p\left(\alpha,\beta,\lambda,w, \nu \mid\textbf{\mbox{X}} \right) $ from the model proposed by $(\ref{DPMM_hierarchical_order_sequences_EW})$ with 
$$\textbf{\mbox{X}} = \{  \boldsymbol{x}_{i,1:n} : x_{i,n-l_i+1} \leq x_{i,n-l_i+2}\leq\ldots\leq x_{i,n}, \text{with } x_{i,j}  \text{ censored to 0 for $ 1\leq j \leq n-l_i$} \},$$
where $\boldsymbol{x_{i}}  $ includes the \it variable length order statistics sequence \rm of  length $l_i$ (uncensored ordered entries), with the remaining $\left(n-l_i\right)$ being censored. There are three steps to obtaining samples from $p\left(\alpha,\beta,\lambda,w, \nu \mid\textbf{\mbox{X}}\right) $ for each MCMC iteration: sampling  the atoms  $\left(\alpha,\beta,\lambda,w\right) $ of the $ DP\left(\nu G_{0} \right)$ for each order statistics sequence;  sampling the cluster-wise atoms for each of the unique clusters (as induced by $DP\left(\nu G_{0} \right)$), and finally, sampling the $\nu$ scale parameter. We initiate by using the Polya urn exposition of a DP \citep{blackwell1973ferguson}  by taking a Gibbs sample of $\boldsymbol{\theta}_{i} = \left(\alpha_i,\beta_i,\lambda_i,w_i\right)$ atoms associated to observation $\boldsymbol{x}_{i}$ using:
\begin{equation} \label{Polya_urn_samplers}
	p\left(\boldsymbol{\theta}_{i}\mid  \boldsymbol{\theta}_{-i}, \nu,\textbf{\mbox{X}}\right) = q_{0}^{*}H_i+\sum_{k=1}^{N^{*}} q_{k}^{*}\delta_{\boldsymbol{\theta}^{*}_{k}}\end{equation}
where $ q_{0}^{*} \propto \nu \int f\left( \boldsymbol{x}_{i}\mid  \boldsymbol{\theta}\right) G_{0} \left(d \boldsymbol{\theta}\right)$ and $ q_{k}^{*} \propto  N^{*}_{k} f\left( \boldsymbol{x}_{i}\mid  \boldsymbol{\theta}_{k}^{*}, \nu\right) $ subject to $ \sum_{k=0}^{N^{*}}q_{k}^{*}  =1 $. Here $f\left( \boldsymbol{x}_{i}\mid  \boldsymbol{\theta}\right) = f_{\left(n\right):\left(n-l_i+1\right)} \left(\boldsymbol{x}_{i} \mid  l_i, \alpha, \beta, \lambda\right)  p\left(l_i \mid  w\right)$, where $f_{\left(n\right):\left(n-l_i+1\right)}$ is specified in $(\ref{EW_joint_VLOSS_density}) $  and the conditional distribution $p\left(l_i \mid  w\right)=\binom{n-1}{l_i-1} w^{\left(l_i-1\right)}\left(1-w\right)^{\left(n-l_i\right)}$.   $H_i$ is the posterior distribution for $\boldsymbol{\theta}$ based on the prior distribution $G_0$ of (\ref{DPMM_Base}) with likelihood $f\left( \boldsymbol{x}_{i}\mid  \boldsymbol{\theta}, \nu\right) $. Here  $\boldsymbol{\theta}_{-i}$ denotes the vectorised atoms of $\boldsymbol{\theta} $ excluding the $i^{th}$ atom $\boldsymbol{\theta}_{i}$,  $\{ \boldsymbol{\theta}_{1}^{*},\ldots,\boldsymbol{\theta}_{N^{*}}^{*} \}$ denotes the unique values of $\boldsymbol{\theta}_{i}$,  $N^{*}$ the number of unique clusters induced by the DP and $ N^{*}_{k}$ the number of points assigned to atom $\boldsymbol{\theta}_{k}^{*}  $. 
As calculating the integral $ q_{0}^{*}$ is intractable, we use algorithm 8 \citep{neal2000markov} to approximate $ q_{0}^{*}$ by a weighted mixture of likelihoods by taking $c$ auxiliary components sampled from the prior distribution $G_0$. Concretely, samples of $\boldsymbol{\theta}_j \overset{iid}{\sim} G_0$ for $j=N^{*}+1,...,N^{*}+c$ are drawn, which then reduces (\ref{Polya_urn_samplers}) to taking a sample from the multinomial distribution given by $ P\left(\boldsymbol{\theta}_i = \boldsymbol{\theta}_{k}^{*} \mid  \boldsymbol{\theta}_{-i}, \boldsymbol{x}_{i}, \boldsymbol{\theta}_{1}^{*},\ldots, \boldsymbol{\theta}_{N^{*}+c}^{*}\right)$, which corresponds to
\begin{displaymath}
  P\left(C_i = k \mid C_{-i}, \boldsymbol{x}_{i}, \boldsymbol{\theta}_{1}^{*},\ldots, \boldsymbol{\theta}_{N^{*}+c}^{*}\right) \propto \left\{
    \begin{array}{lr}
      \frac{N^{*}_{k}}{N-1+\nu}f\left( \boldsymbol{x}_{i}\mid  \boldsymbol{\theta}_{k}^{*} \right) \;\;1 \leq k \leq N^{*}\\
      \frac{\nu/c}{N-	1+\nu}f\left( \boldsymbol{x}_{i}\mid  \boldsymbol{\theta}_{k}^{*} \right)  \;\;  N^{*}< k \leq N^{*}+c
    \end{array}
  \right.
\end{displaymath}
The number auxiliary components $c$ chosen determines the level to which  $ q_{0}^{*}$ is approximated to. 
Finally, the  $\boldsymbol{\theta}_{k}$ atoms are then updated for each of the unique clusters $ k=1,\ldots,N^{*}$ to avoid inefficiencies associated with having to pass through extremely low probability states to get to a higher probability states. This is achieved by taking a single sample from the posterior $ p\left(\boldsymbol{\theta}_{k} \mid \nu, \boldsymbol{x}_{\{i:\,C_i=k\}}\right) $ for each $ k=1,\ldots,N^{*}$. As taking exact samples from  $ p\left(\boldsymbol{\theta}_{k} \mid \nu, \boldsymbol{x}_{\{i:\,C_i=k\}}\right) $ is intractable for our choice of kernel (\ref{EW_joint_VLOSS_density}) and prior $G_0$ (\ref{DPMM_Base}),  the Metropolis Hastings algorithm is used to sample from $ p\left(\boldsymbol{\theta}_{k} \mid \nu, \boldsymbol{x}_{\{i:\,C_i=k\}}\right) $ for $ k=1,\ldots,N^{*}$. This involves taking sufficient burn-in samples until  convergence to the stationary posterior distribution is satisfactory, at which point $\boldsymbol{\theta}_{k}$ is then taken as the last sample from the Metropolis Hastings procedure. Finally,   $ \nu$ is updated in line with \cite{escobar1995bayesian} auxiliary variables approach.  Further details of the posterior inference steps is included in Appendix $\ref{sec:AppendixA}$.

\section{\label{sec:DataExamples}Data examples}
  
We now illustrate how our methodology works in practice by performing two simulation studies, before proceeding to a real-world retail analytics dataset. In the first example (subsection \ref{subsec:SimulatedData1}) we generate data from our model. In the second example (subsection \ref{subsec:SimulatedData2}) we generate data using a Gamma distribution as a kernel (rather than EW). We then fit our model to both datasets using vague priors.

%
\subsection{Simulated data 1}\label{subsec:SimulatedData1}
We generate data using parameters for the mixtures of (\ref{DPMM_hierarchical_order_sequences_Sim1})  which demonstrate the various behaviours that variable length order statistics sequences from an EW kernel can exhibit, namely a mixture of light and heavy tails with varying rates of order statistics terms ${x}_{i,\left(20 \right)}$ convergence to 0, lengths, different decay rates and varying modal behaviours.
Specifically, we draw 500 samples from the following DPMM of \it variable length order statistics sequences \rm of (\ref{DPMM_hierarchical_order_sequences_EW}):
\begin{equation} \label{DPMM_hierarchical_order_sequences_Sim1}
  \begin{split} 
 G & =   0.4\delta_{\boldsymbol{\theta}_1}+0.35\delta_{\boldsymbol{\theta}_2}+0.25\delta_{\boldsymbol{\theta}_3} \\
 \left(\alpha_i,\beta_i,\lambda_i,w_i\right) \mid G & \sim G,\;i=1,\ldots,500, \\
 l_i & \sim 1+Binomial\left(19,  w_i \right),\;i=1,\ldots,500, \\
x_{i,j} &\sim EW\left(\alpha_i,\beta_i,\lambda_i\right),\;j=1,\ldots,20,
  \end{split}
\end{equation}
where $\boldsymbol{\theta} = $ $\left(\alpha,\beta,\lambda,w \right)$, with  $\boldsymbol{\theta}^*_1 =\left(0.15,0.8, 0.91,0.65 \right)$, $\boldsymbol{\theta}_2^*= \left( 2.5, 3.3, 0.35,0.75 \right)$, $\boldsymbol{\theta}^*_3= \left( 0.64, 1.7, 0.4,0.9 \right)$, and within each observation vector $i$ with all but the top $l_i$ entries being censored.

\subsection{Simulated data 2}\label{subsec:SimulatedData2}
 This simulated example differs from the former simulation study in that the data is simulated from a mixture of gamma distributions rather than a mixture EW distributions. The purpose of fitting our model to a mixture of Gamma distributions instead of a mixture of EW distributions is to test the inference in a less optimistic setting and establish whether the EW kernel is sufficiently flexible to capture the decay of order statistics sequences from a set of mixtures that are not a mixture of EW distributions. The mixture components $\boldsymbol{\theta}^*_1, \boldsymbol{\theta}^*_2, \boldsymbol{\theta}^*_3 $ are selected to produce simulated mixtures that imitate the mixtures of (\ref{DPMM_hierarchical_order_sequences_Sim1}).

We generate 500 samples from the following DPMM of \it variable length order statistics sequences \rm from the following mixture model
\begin{equation} \label{DPMM_hierarchical_order_sequences_Sim2}
  \begin{split} 
 G & =   0.4\delta_{\boldsymbol{\theta}_1}+0.35\delta_{\boldsymbol{\theta}_2}+0.25\delta_{\boldsymbol{\theta}_3} \\
 \left(\alpha_i,\beta_i,w_i\right) \mid G & \sim G,\;i=1,\ldots,500, \\
 l_i & \sim 1+Binomial\left(19,  w_i \right),\;i=1,\ldots,500, \\
x_{i,j} &\sim Gamma\left(\alpha_i,\beta_i\right),\;j=1,\ldots,20,
  \end{split}
\end{equation}
where $\boldsymbol{\theta} = $ $\left(\alpha,\beta,w \right)$, with  $\boldsymbol{\theta}^*_1 =\left(0.15,0.5,0.65 \right)$, $\boldsymbol{\theta}_2^*= \left( 1.7, 1,0.75 \right)$, $\boldsymbol{\theta}^*_3= \left( 32, 10,0.9 \right)$, and within each observation vector $i$ with all but the top $l_i$ entries being censored.  

\subsection{Prior distributions and posterior sampling}\label{subsubsec:MCMCoutputSimulatedData}
We fit our EW mixture model using the following vague priors  for $\left(\alpha,\beta,\lambda,w\right)$ and $ \nu$ (DP scale parameter): 
\begin{align*}
  \left(\alpha,\beta,\lambda,w\right) & \sim  Gamma\left(1,0.1\right) \times Gamma\left(1,0.1\right)\times Gamma\left(1,0.1\right)  \times Beta\left(1,1\right) \\
  \nu & \sim Gamma\left(1,1\right)   
\end{align*}
respectively. The priors for $\alpha$, $\beta$ and $\lambda$ imply a mean of 10 and variance 100, a rather vague choice centred away from the true values. The $Beta$ prior for $w$ corresponds to a uniform distribution, assuming no prior information about the number of non-censored entries. 

 We use the steps outlined in Section \ref{sec:posterior_inference} for parameter inference and perform 10000 MCMC iterations with 200 burn-in, and thin every 10 samples after the subsequent burn-in samples are discarded. We present the MCMC output based on the inference methodology of Section \ref{sec:posterior_inference} on the simulated data of (\ref{subsec:SimulatedData1}) and (\ref{subsec:SimulatedData2}). Since individual clusters are not identifiable (up to permutations), an additional identifiability criterion is required in order to perform cluster-wise inference. 
 We implement algorithm 2 of \cite{lau2007bayesian} on the posterior samples to derive an `optimal' partition. \cite{lau2007bayesian} select the partition of the observations into clusters $\boldsymbol{C}^*$ (with some permutation) which minimise a linear loss function of the posterior expected loss of the posterior marginal coincidence probabilities. This is equivalent to maximising
 \begin{equation} \label{LG_max}
l\left(\boldsymbol{C}^* \rm, K\right) = \sum_{\left(i,j\right) \in M}  \mathbb{I} \left[{C}^*_i = {C}^*_j \right]\left(\rho_{ij}-K\right)
\end{equation} 
where $ M=\{\left(i,j\right):i<j;i,j \in\{1,\ldots,N\}\}$, $ K = \frac{b}{a+b} \in \left[0,1 \right] $ where $b$ is the penalty of misclassifying two points into different clusters (when they should be) and $a$ the penalty of misclassifying two points being the same cluster (when they shouldn't be), $\boldsymbol{C}$ is a given clustering of the observations (up to permutation) and $\rho_{ij}  $ is the posterior coincidence probability between points $i,j$.  This optimal partition defines cluster assignments $C^*_i$ for each observation $\boldsymbol{x}_i$ and is used in future sections to compute cluster-wise point estimates of various quantities of interest.  As (\ref{LG_max}) needs to be maximised over $K$, we maximise (\ref{LG_max}) over each $K \in \{0.1,0.2,0.3,0.4,0.5,0.6,0.7,0.8,0.9\}$ and select $K$ and $\boldsymbol{C}^*$ that give the maximal value. 

The partition  $\boldsymbol{C}^*$ obtained through the above algorithm determines an optimal number of clusters and an optimal allocation of each observation into a cluster. Posterior distributions of clusterwise parameters are then obtained by sweeping through MCMC samples and, at each iteration, averaging over all parameter values associated with each observation within a cluster of $\boldsymbol{C}^*$. 

\begin{figure}
  \centering
\subfigure[]{
\includegraphics[trim = 0cm 0cm 0.0cm 0cm, clip, width=0.31\textwidth]{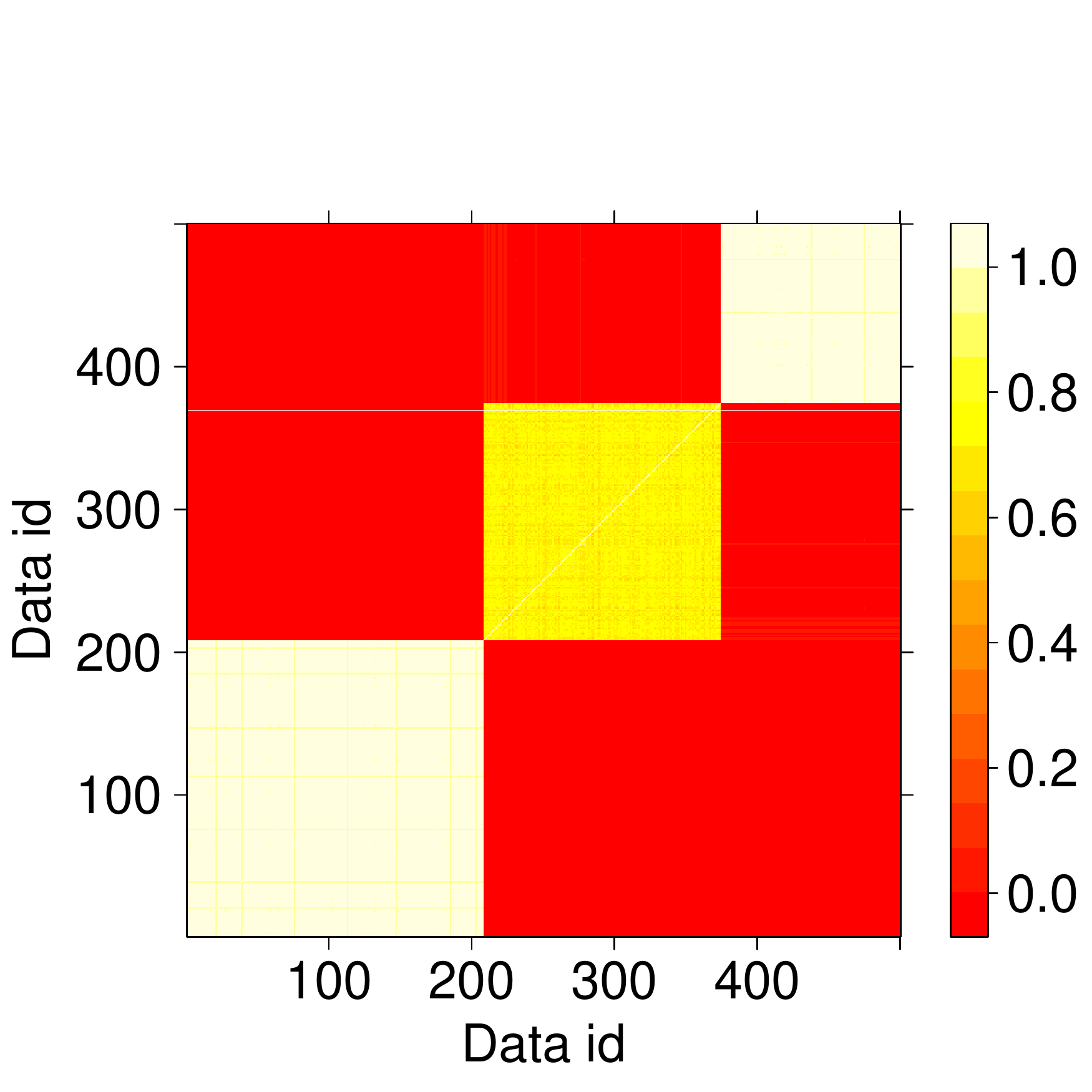}\label{fig:f2}
}
  \subfigure[]{
\includegraphics[ width=0.31\textwidth]{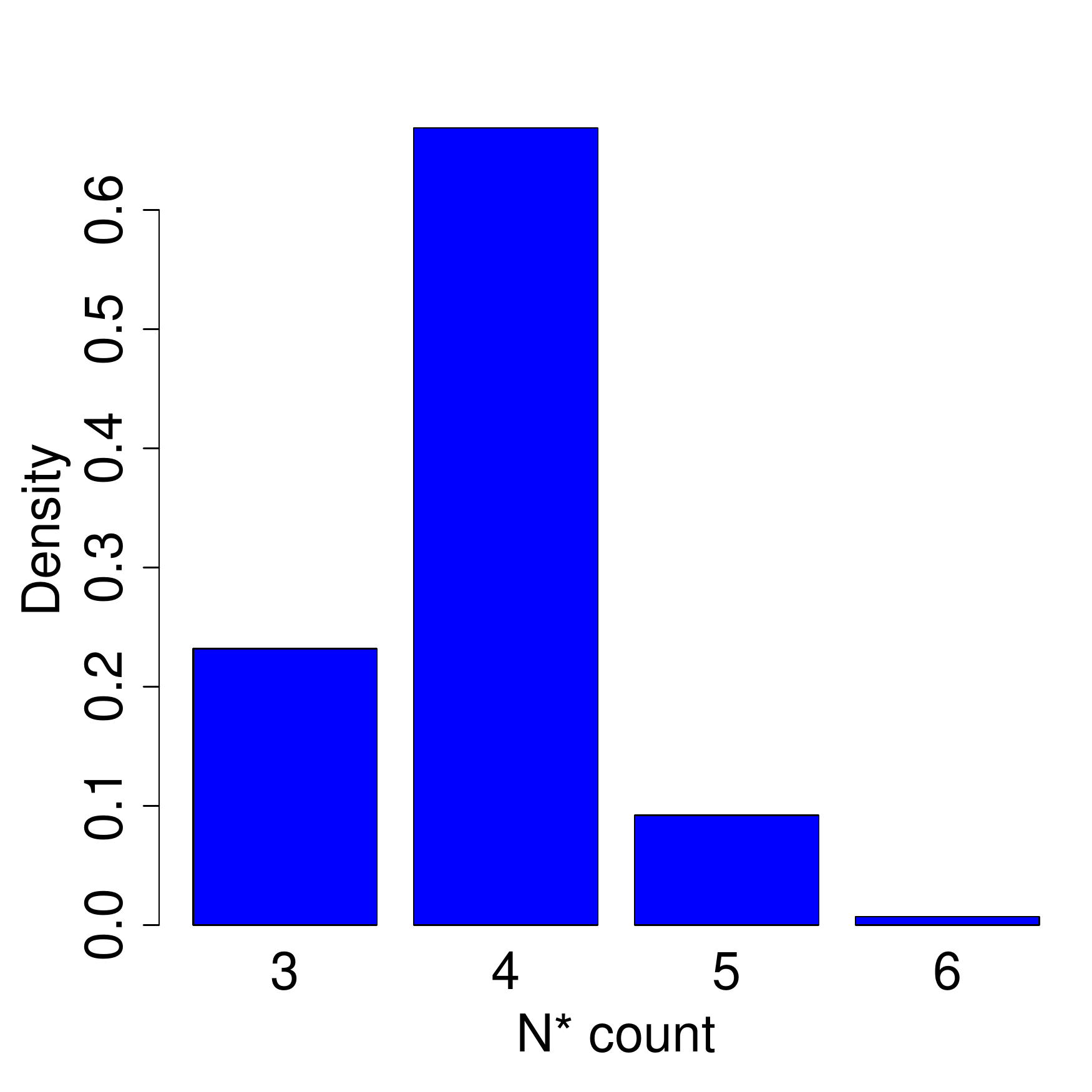}\label{fig:f1}
}
   \subfigure[]
    {
      \includegraphics[width=0.31\textwidth]{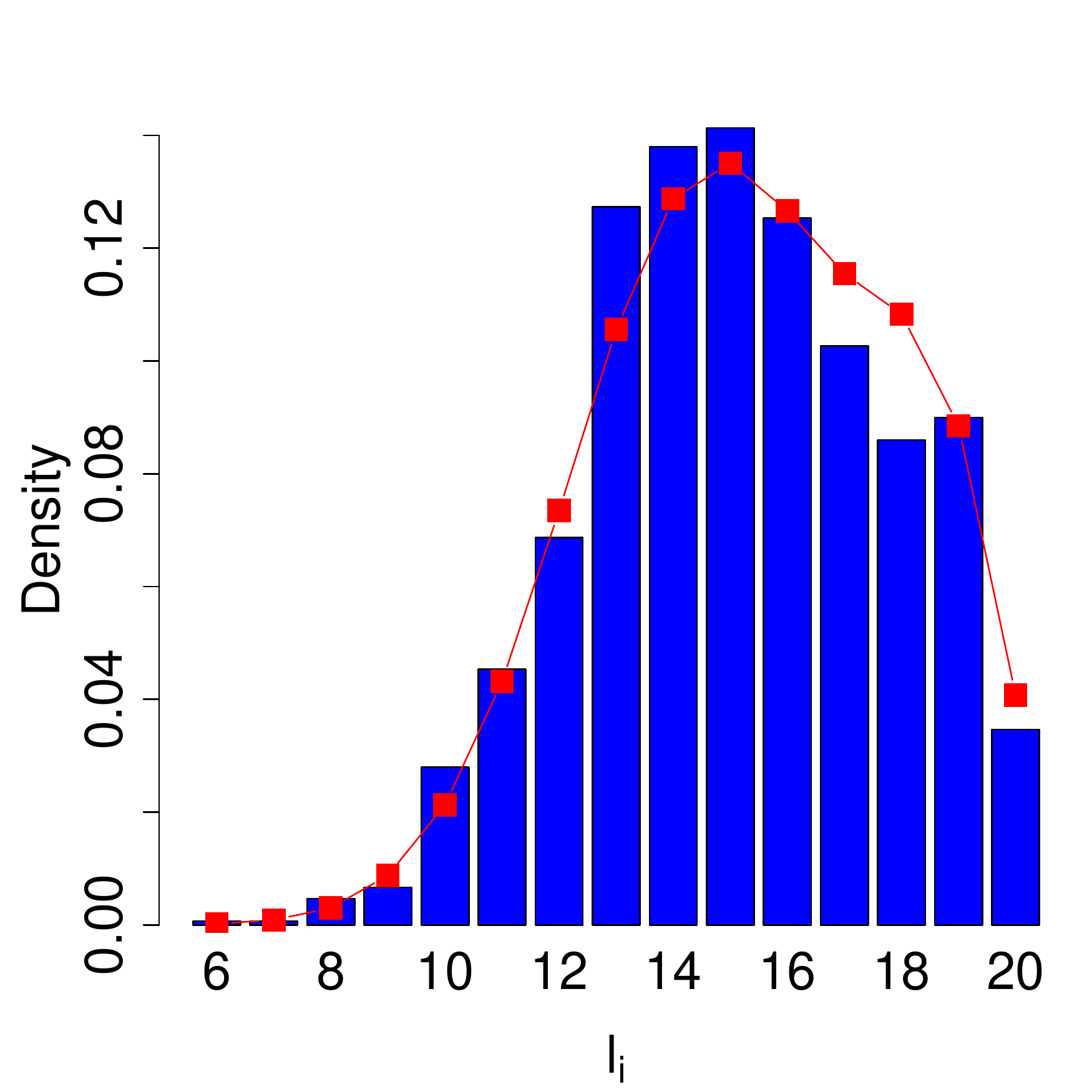}
    }
    \\
   \subfigure[]
    {
      \includegraphics[width=0.31\textwidth]{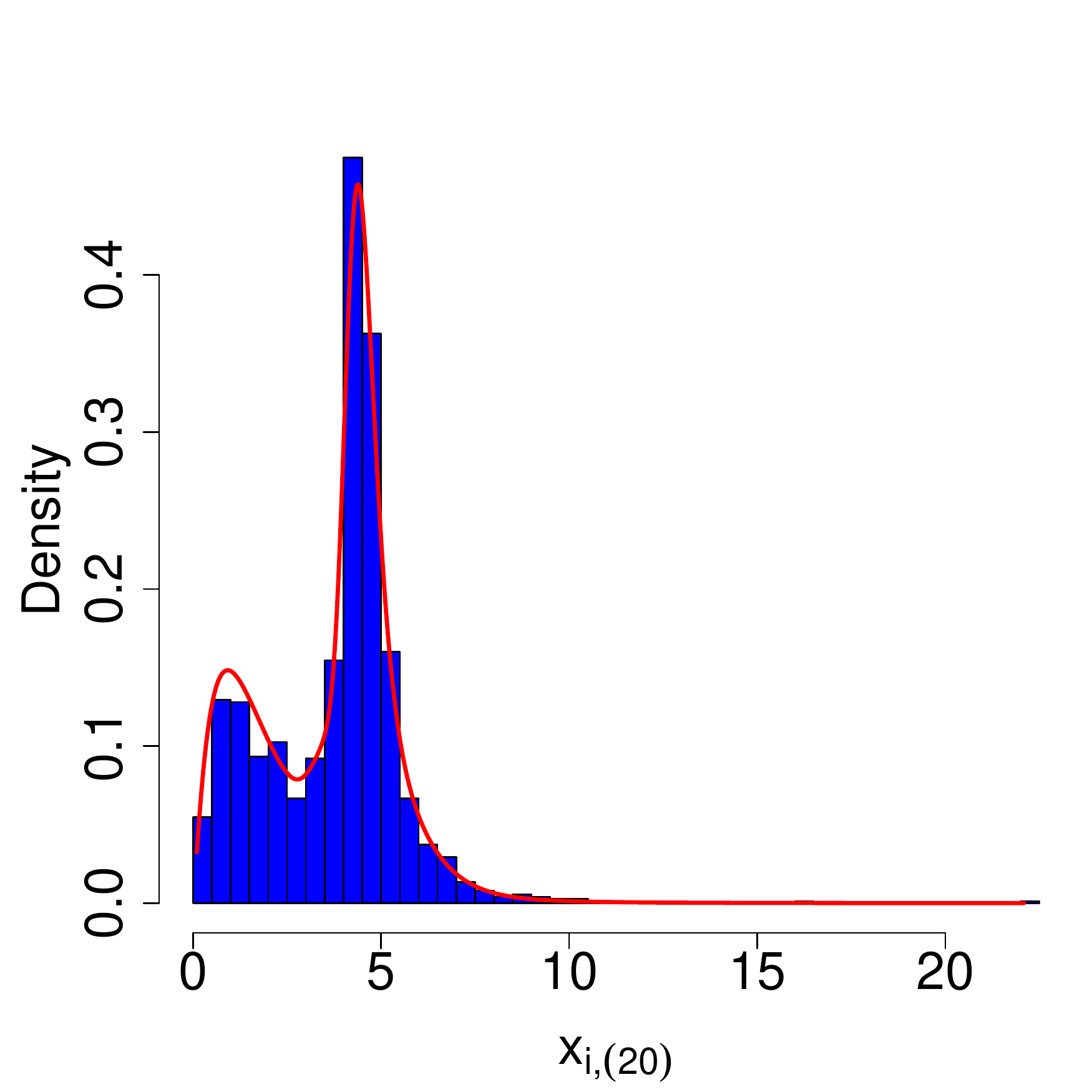}
    }
   \subfigure[]
   {
      \includegraphics[width=0.31\textwidth]{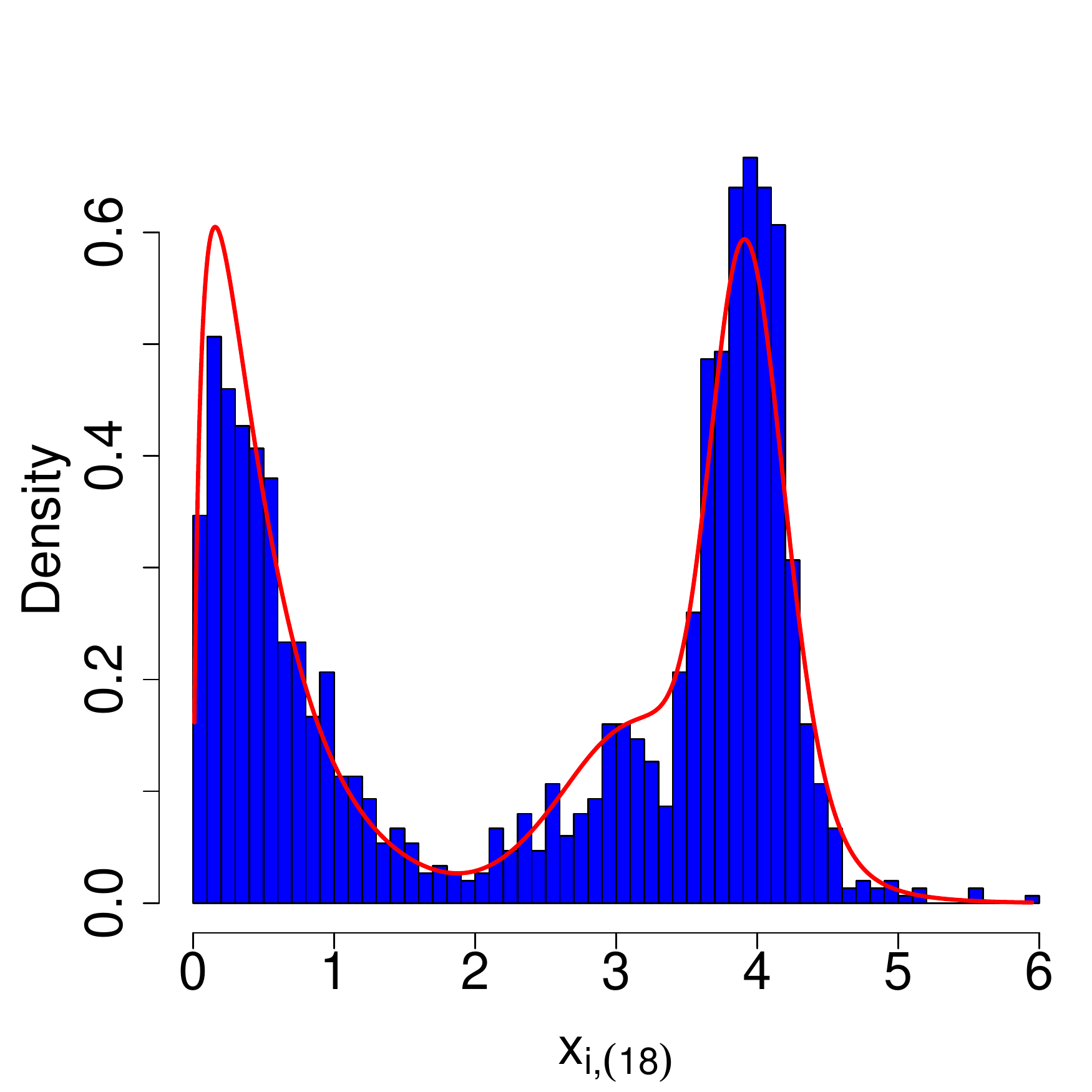}
    }
   \subfigure[]
    {
      \includegraphics[width=0.31\textwidth]{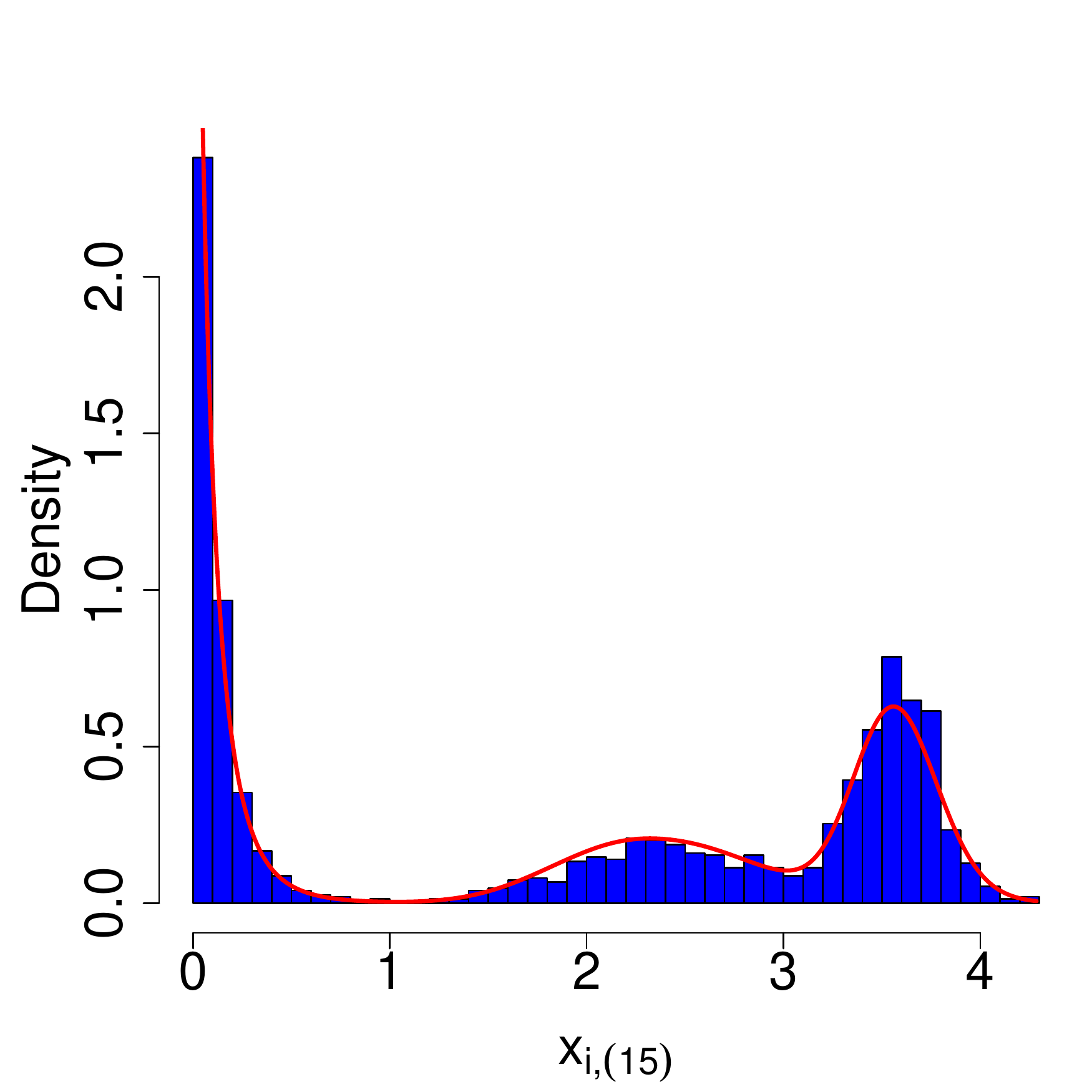}
    }
     \caption{Posterior probability cluster co-membership probability heatmap, histograms and density estimates for  $N^*$, $l$  and a few order statistics of simulated data (\ref{DPMM_hierarchical_order_sequences_Sim1}). }\label{plot:MCMCoutput_simulated1}
\end{figure}

\begin{figure}[h]
  \centering
\subfigure[]{
\includegraphics[trim = 0cm 0cm 0.0cm 0cm, clip, width=0.31\textwidth]{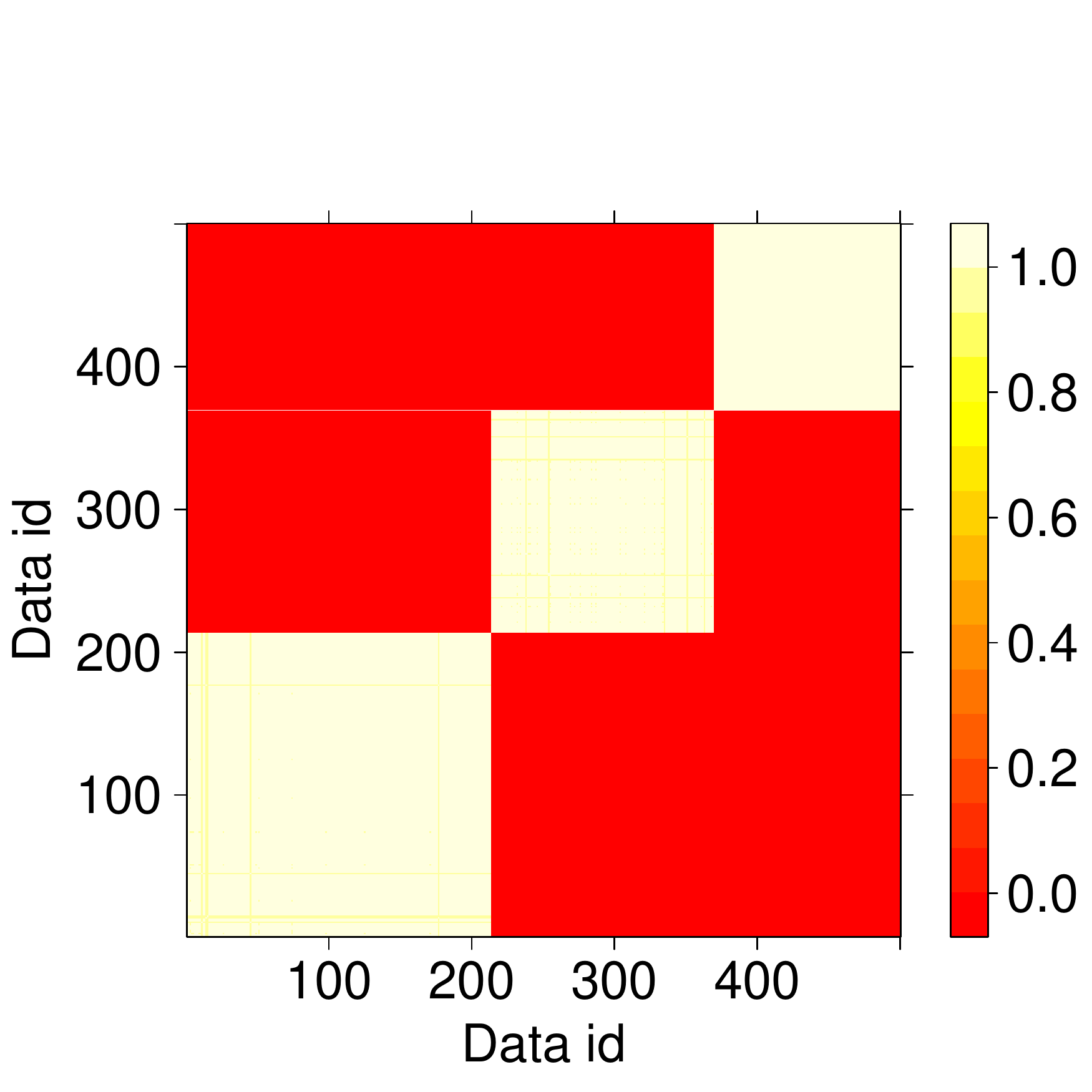}\label{fig:f2}
}
  \subfigure[]{
\includegraphics[ width=0.31\textwidth]{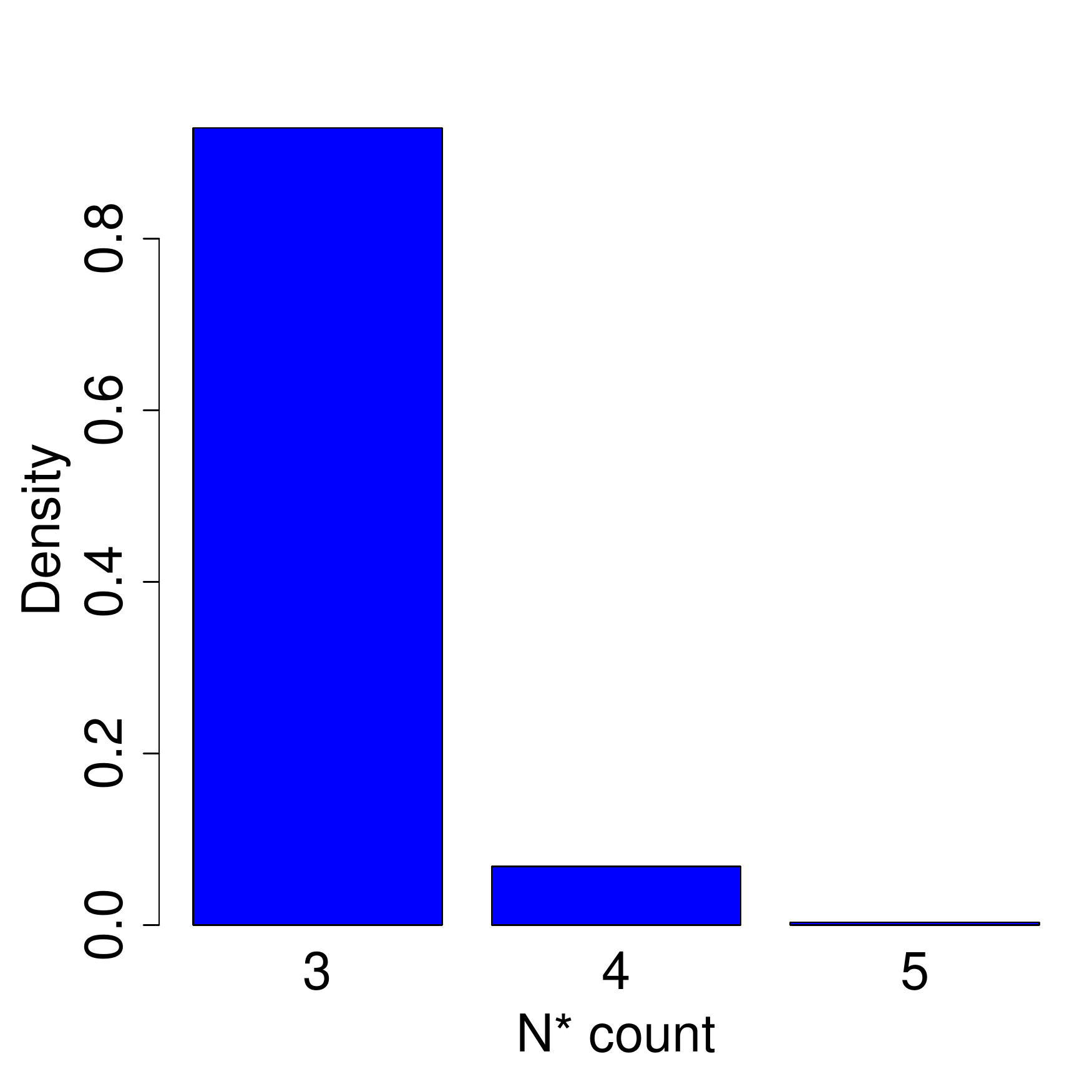}\label{fig:f1}
}
   \subfigure[]
    {
      \includegraphics[width=0.31\textwidth]{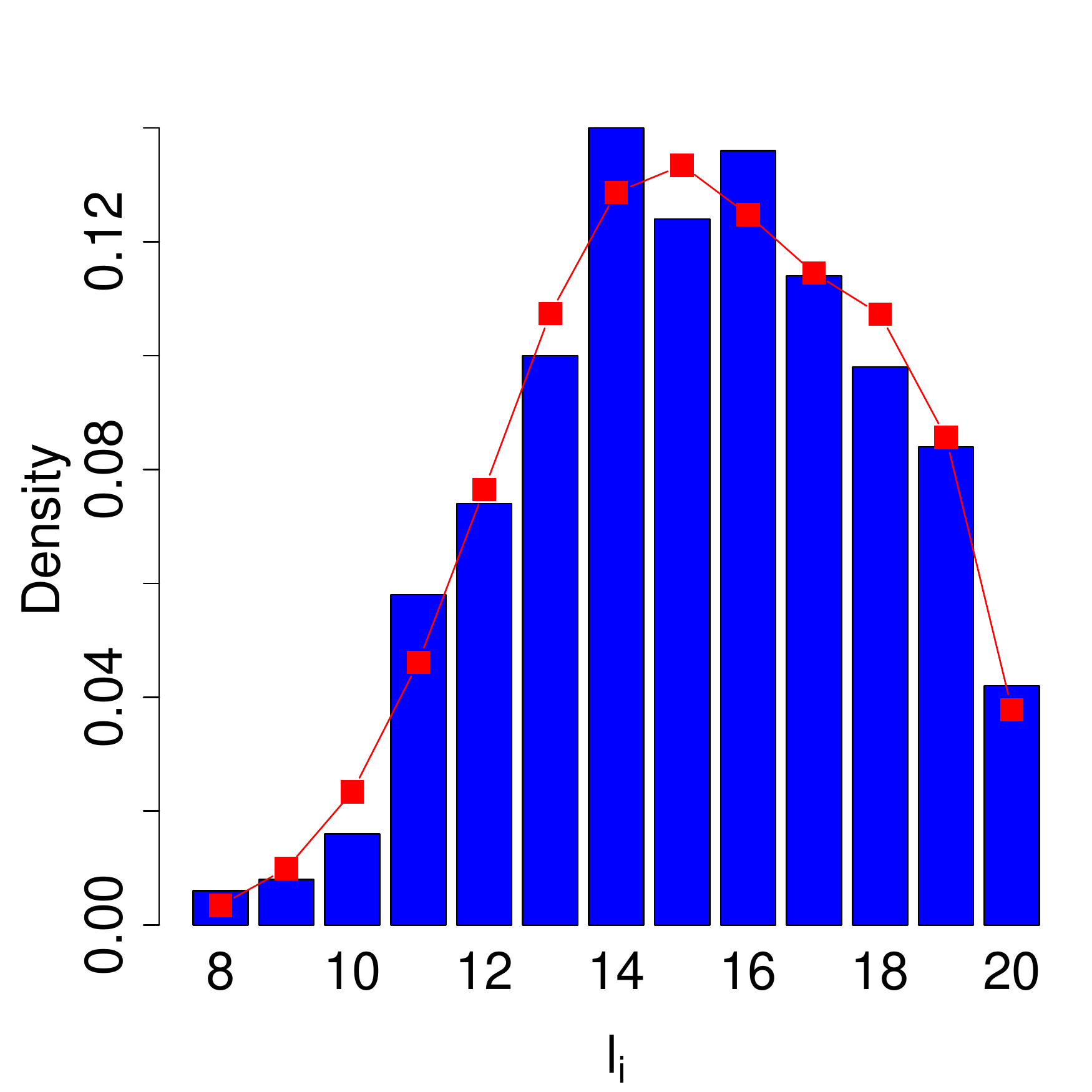}
    }
    \\
   \subfigure[]
    {
      \includegraphics[width=0.31\textwidth]{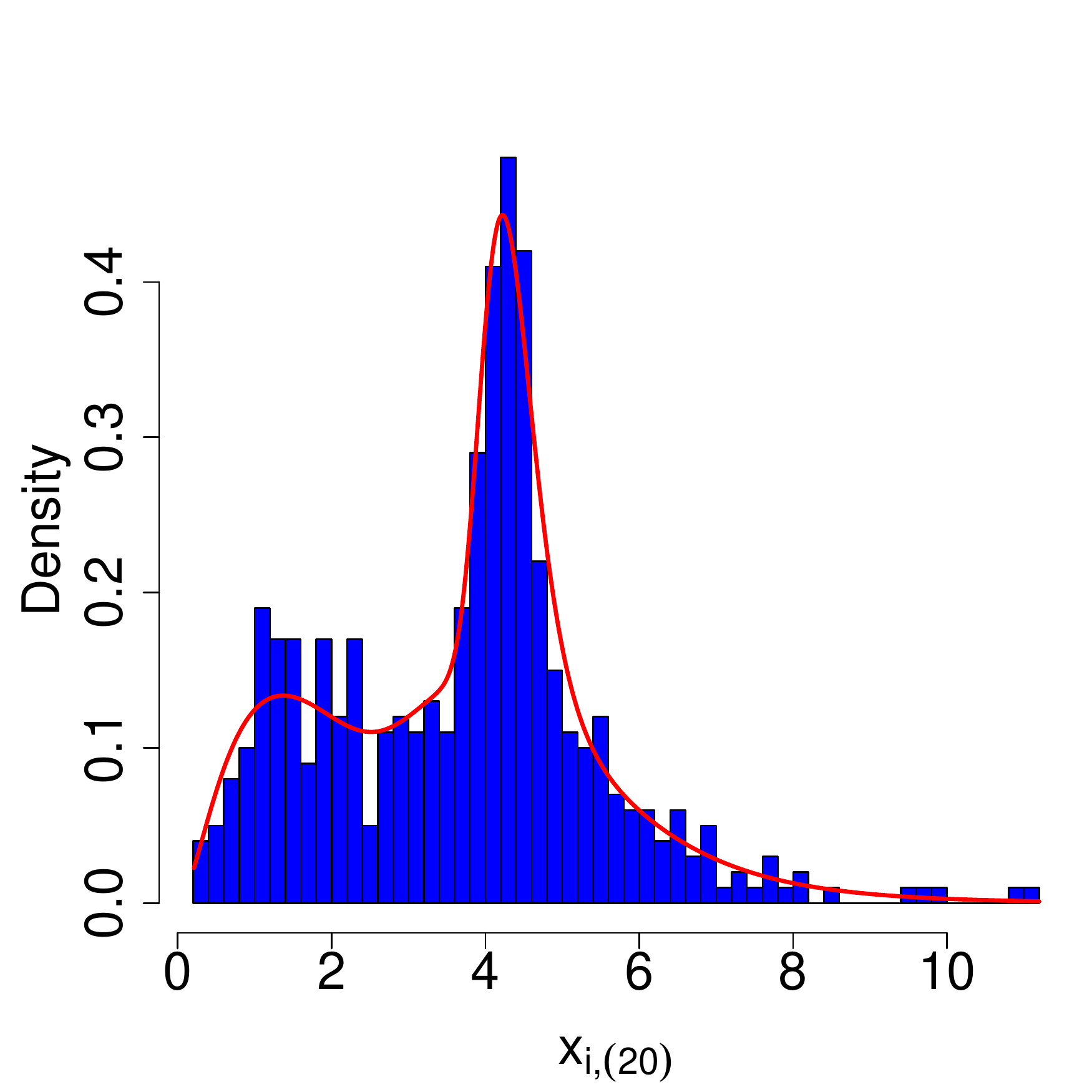}
    }
   \subfigure[]
   {
      \includegraphics[width=0.31\textwidth]{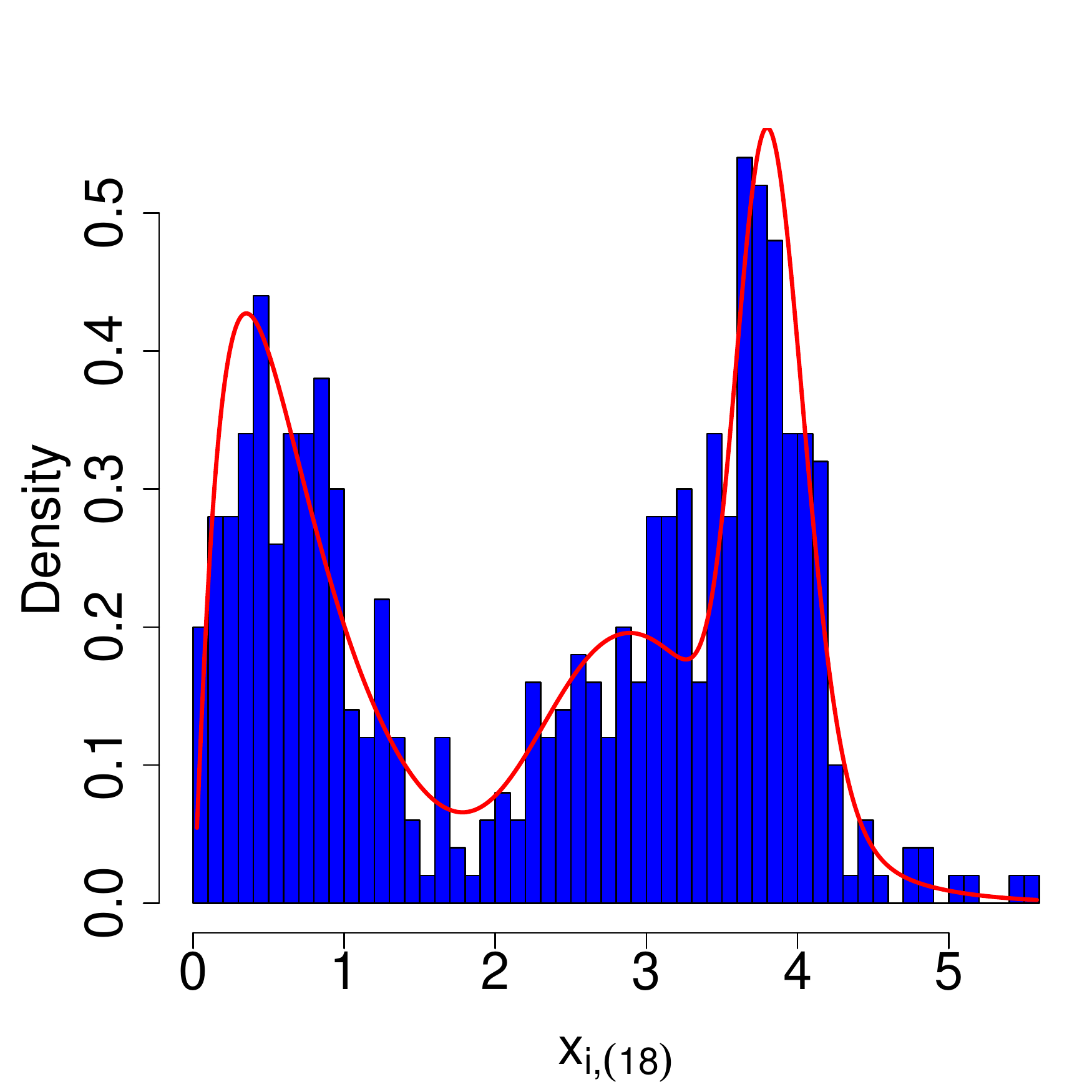}
    }
   \subfigure[]
    {
      \includegraphics[width=0.31\textwidth]{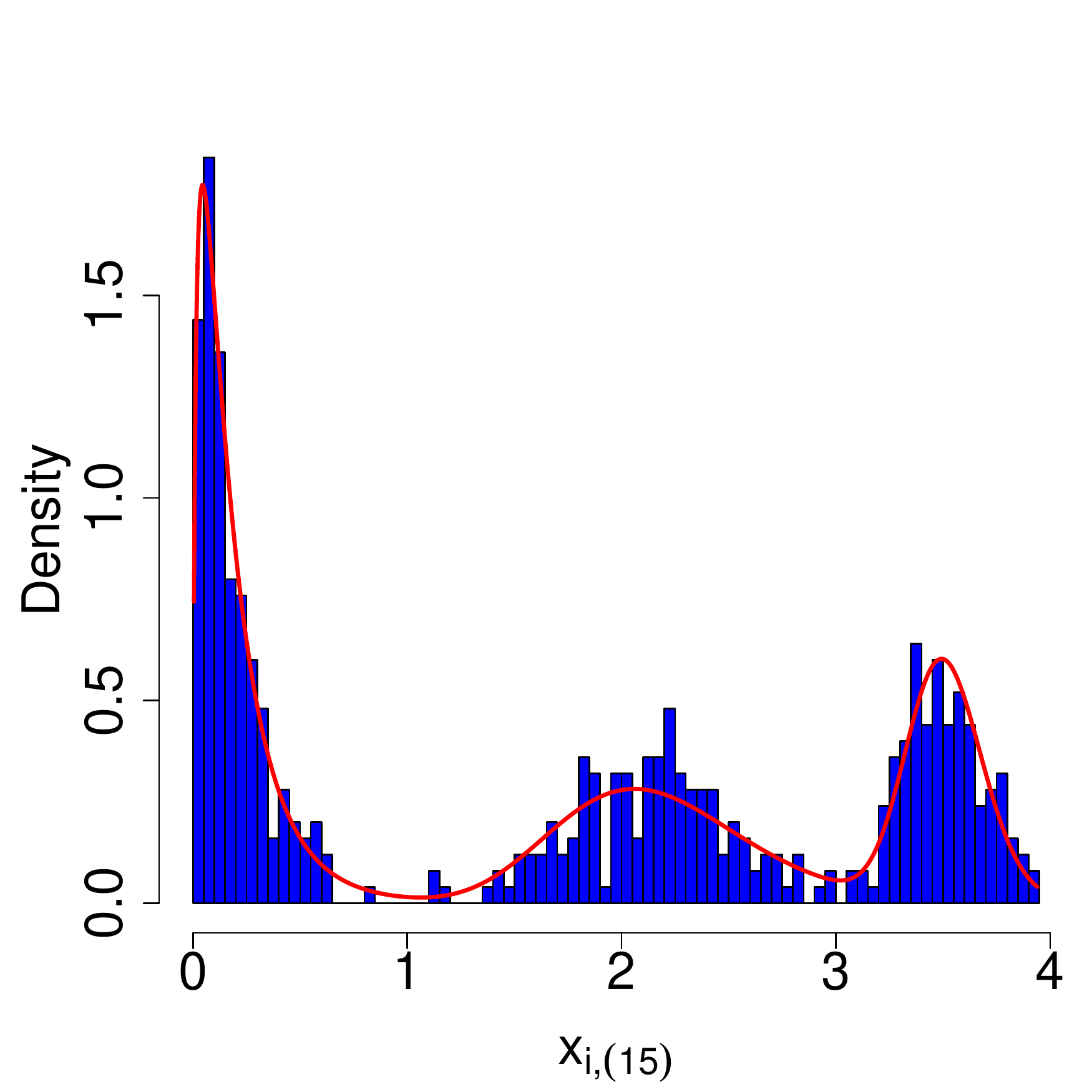}
    }
     \caption{Posterior probability cluster co-membership probability heatmap, histograms and density estimates for  $N^*$, $l$  and a few order statistics of simulated data (\ref{DPMM_hierarchical_order_sequences_Sim2}). These density plots demonstrate the EW kernel successfully describes the mixture of decay sequences despite the data being generated from a mixture of Gamma distributions.}\label{plot:MCMCoutput_simulated2}
\end{figure}

Figures $\ref{plot:MCMCoutput_simulated1}$ and $\ref{plot:MCMCoutput_simulated2}$ present plots on the MCMC output which includes the histogram of the number of occupied clusters $N^{*}$, a heatmap of the posterior marginal coincidence probabilities (cluster co-membership) and density estimates of the order statistics sequences $\boldsymbol{x}_{i}$. The Maximum-A-Posteriori estimate for the number of clusters is indeed 3, with a clear separation of the observations into three clusters, and the marginal density estimates closely match the corresponding histograms of the data.

Table \ref{tab:simulated_table_results1} summarises our MCMC output for the simulated study (\ref{subsec:SimulatedData1})  using the clusters defined by $\boldsymbol{C}^*$. The estimates are very close to the true parameter values, which are also contained within the $95\%$ credibility intervals indicating the inference is working effectively.

\begin{table}
\caption{\label{tab:simulated_table_results1} Posterior means and (2.5\%,97.5\%) credible intervals for the parameters $\left(\alpha, \beta, \lambda, w\right)$ of each cluster from simulated data (\ref{DPMM_hierarchical_order_sequences_Sim1}). The top and bottom rows show the true parameters and number of observations assigned to each cluster respectively}
  \begin{tabular}{c c c c}
    \hline\hline
    & Cluster 1 & Cluster 2 & Cluster 3 \\ [0.5ex] 
     $\left(\alpha, \beta, \lambda, w\right)$ = &  (0.15,0.80,0.91,0.65) &(2.5,3.3,0.35,0.75) & (0.64,1.7,0.40,0.90) \\
    \hline\hline
     $\alpha$ & 0.14  (0.12, 0.17) &   5.2 (1.4, 19)    &    0.61  (0.52, 0.73)\\
    $\beta$ & 0.82 (0.71,  0.94) &      2.9 (1.7, 4.5)    &    1.7 (1.6, 1.9) \\
   $\lambda $ & 0.99 (0.80, 1.2) &    0.41 (0.30, 0.64)   &  0.40 (0.36, 0.44)   \\
  $w $   & 0.66 (0.65, 0.68)          & 0.75 (0.68, 0.81)     & 0.91 (0.89, 0.92)\\
        \hline 
    $N$ & 208 & 166 & 126 \\ 
        \hline
  \end{tabular}
\end{table}

\subsection{Retail analytics dataset}\label{sec:Retail_analytics_dataset}
\begin{figure}[h]
  \centering
\includegraphics[ width=0.31\textwidth]{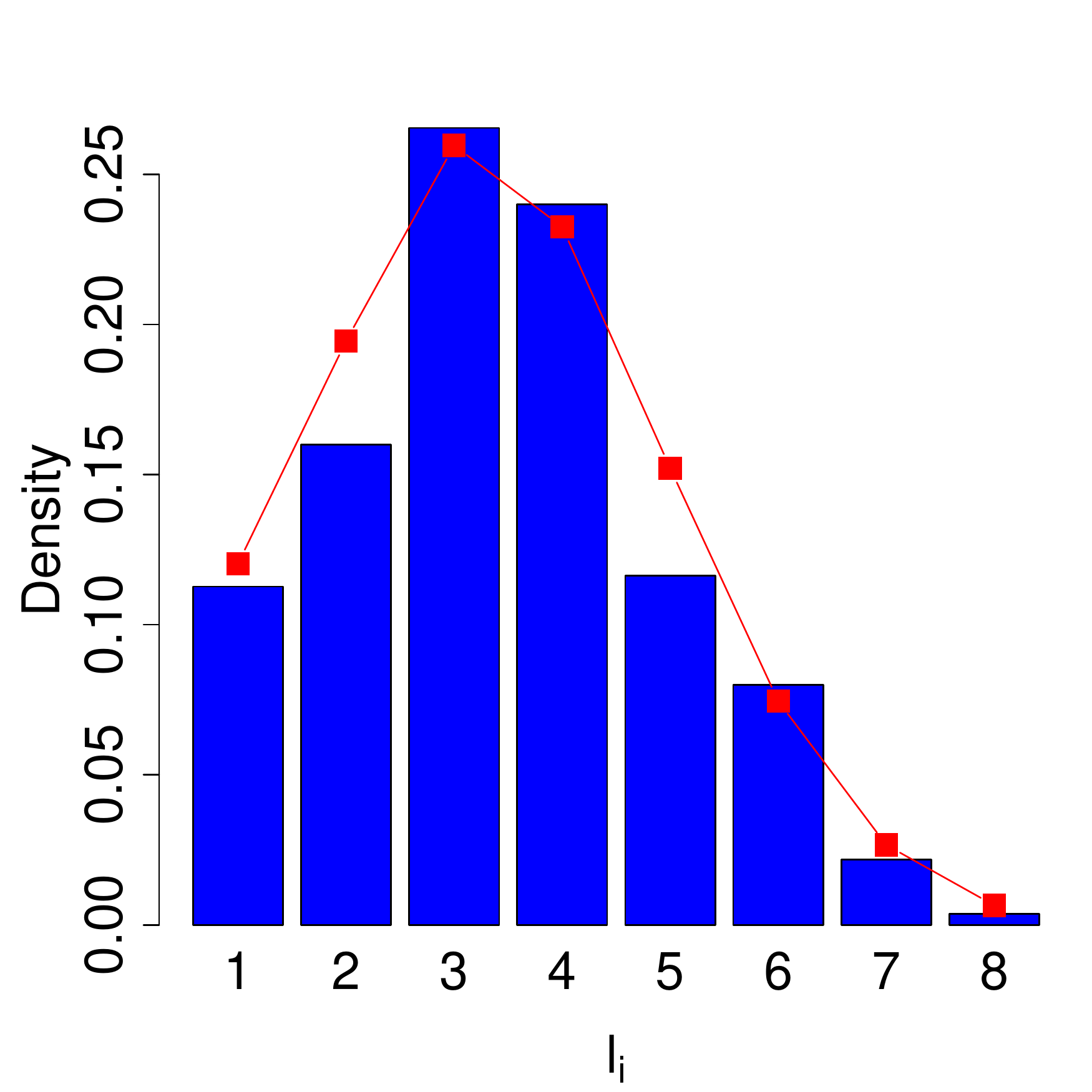}
\includegraphics[ width=0.31\textwidth]{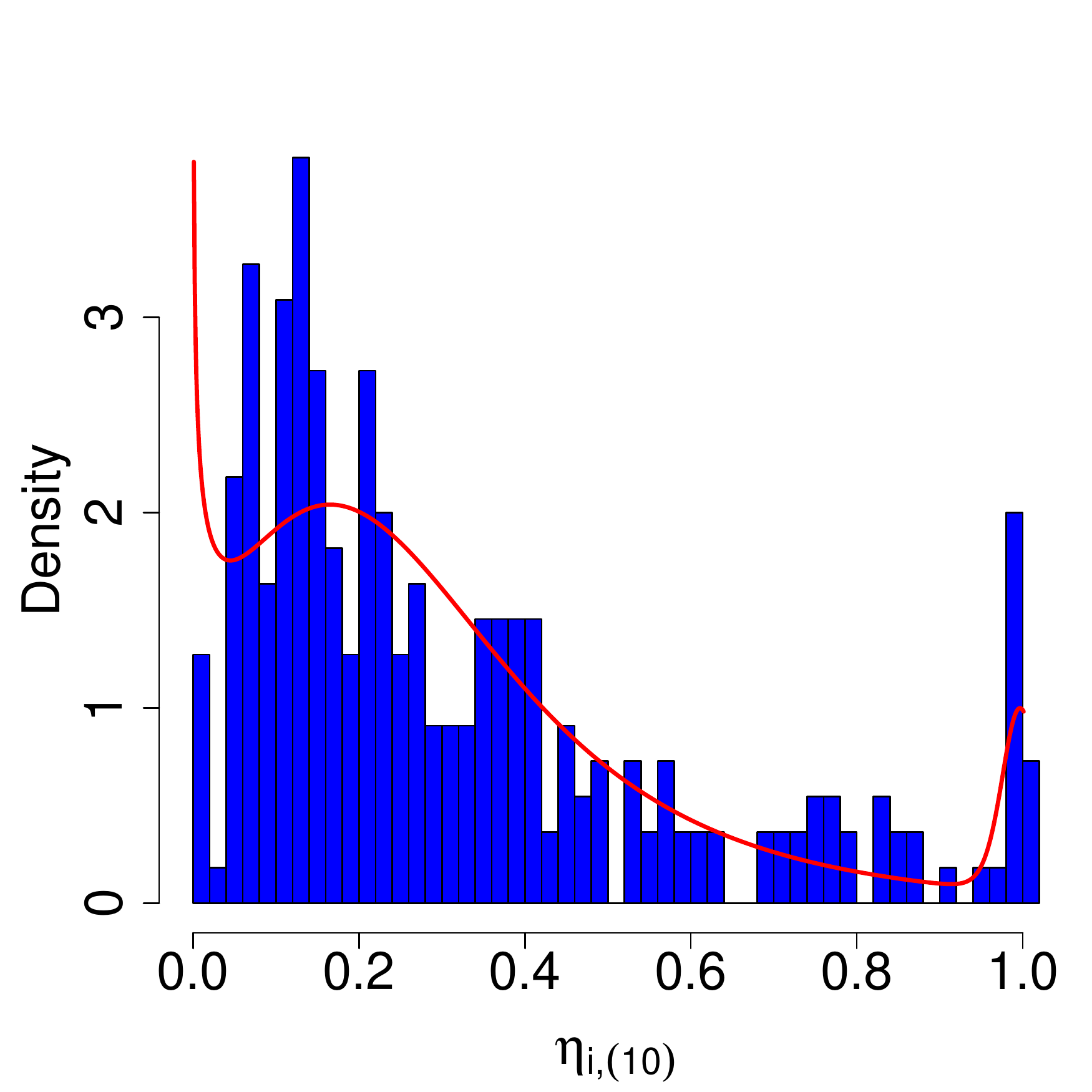}
\includegraphics[ width=0.31\textwidth]{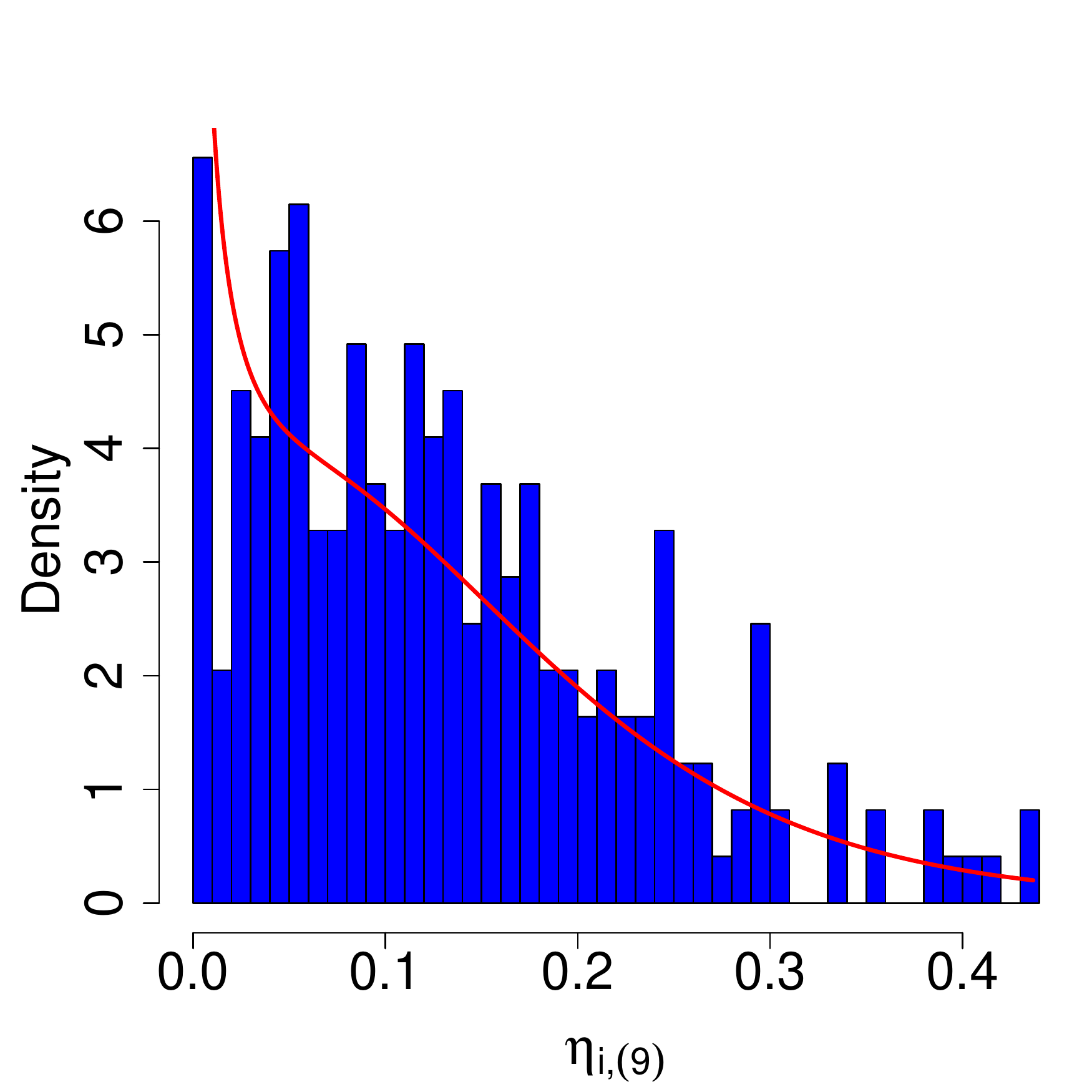}
 \caption{\label{fig:RA_data}Histograms of the number of observed entries in each cross-elasticity vector, as well as the top two entries of the cross-elasticity vectors, with corresponding density estimates from our model. The censored entries (corresponding to 0 elasticities) have been omitted from the histograms.}
\end{figure}

We apply our method of order statistics clustering to a retail analytics dataset from a leading UK supermarket chain. Access to the anonymised data was provided by dunnhumby. The dataset consists of the cross elasticities for a category of supermarket products of the format described in Section \ref{sec:Motivation}.
$$\textbf{\mbox{X}} = \{  \boldsymbol{\eta}_{i,1:n}  : \eta_{i,\left(n-l_i+1\right)}\leq \ldots\leq \eta_{i,\left(n\right)}, \text{with } \eta_{i,\left(j\right)} =  \text{censored to 0 for $ 1\leq j \leq n-l_i$} \},  $$
where we have observed only the top $l_i$ order statistics of each cross-elasticity vector $\boldsymbol{\eta}_i$.
To allow for straightforward interpretation we focus on the snacks category which consists of $N=275$ (out of thousands) products, so that observations consist of $N=275$  vectors of cross elasticity coefficients. For this study, a maximum of $n=10$ competitors is considered a priori to reflect a product's most significant competitors.
The snack category consists of the following product line break-down: 22.5\% traditional flavoured crisps (salted, cheese and onion, salt and vinegar), 33.1\% exotic flavoured crisps  (crisps excluding traditional flavours), 8.73\% tortillas, 8.00\% popcorn, 7.64\% nuts, 4.73\% dips, 2.18\% pretzels and 13.1\% other peripheral quick snack products. 
Figure \ref{fig:RA_data} shows summary plots for the snacks category in this study, although other categories will show different behaviour. Specifically, we show histograms of the lengths $l_i$ of $\boldsymbol{\eta}_i$ as well as the top two terms of the sequences. The histogram of the top order statistics demonstrates spikes centred around 0.0 and 1.0, suggesting possible multi-modality.

\subsubsection{Omitted competitors \& aggregate competition}\label{subsubsec:OCandMAE_defs}
We introduce two statistics relevant to  the retail analytics setting; \it omitted competitors \rm and \it mean aggregate competition\rm. These notions have key interpretations in the retails analytics context and will allow us to assess model fit.
\newline

\textbf{Definition 1}: \emph{Omitted competitors}

As discussed in Section \ref{sec:Motivation}, censoring of lower order statistics in the cross-elasticity vector occurs through penalised regression. However, it is possible for potentially important competitor products to have been inadvertently omitted from the regression equation, meaning that the cross-elasticity vector should have included additional uncensored entries. 
%
%
%
The objective of the \it omitted competitors \rm (OC) statistic is to assess whether the truncation has occurred prematurely by predicting the subsequent term of the observed order statistics sequence (i.e.~$\eta_{i,\left(n-l_i \right)}$ of $\boldsymbol{\eta}_i$) and assessing whether this predicted value is sufficiently large. 
 Concretely, we say an elasticity vector contains omitted competitors if its \it variable length order statistics sequence \rm satisfies
$$ \textrm{OC} = \mathbb{E}_{\tilde{l},\tilde{{\eta}}_{\left(n-\tilde{l} \right)} } \left[\tilde{{\eta}}_{\left(n-\tilde{l} \right)}\mid \alpha, \beta, \lambda, w \right] \geq \epsilon, $$
for some truncation constant $ \epsilon >0$ and where $\tilde{\eta}_{(n-\tilde{l})}$ represents the random quantity of the $(n-\tilde{l})th$ order statistic of $n$ i.i.d.~$EW\left(\alpha, \beta, \lambda \right)$ samples with $ \tilde{l} \sim 1+Binomial\left( n-1, w\right)$. In other words, $\tilde{\eta}$ has the same distribution as $\eta$, but without any censoring. Thus the OC statistic represents the expected value of the $1^{st}$ censored term of a cross-elasticity vector $\tilde{\boldsymbol{\eta}}$, were we to have observed it. The value of $\epsilon$ should be chosen to represent a `small value' within the modelling context. We set $\epsilon =0.05 $ as a sensible value to deem truncation (and will be fixed for our subsequent analysis) as it implies that if log price deviations of the next competitor is expected to account for more than 5\% of equivalent log prices changes of the product's own cross-elasticity coefficient ${\varphi}_{i}$, we conclude this as a significant omission in the sales model. 
One of the benefits of interpreting the cross-elasticities as  \it variable length order statistic sequences \rm is the utility it provides with respect to defining OC statistic by casting censored observations into a missing data framework. The OC statistic crucially relies on being able to make a prediction of the subsequent value of a cross-elasticity vector were it to be observed. The \it variable length order statistic sequence \rm model, by capturing the sequential decay of these decreasing sequences, allows inferences on subsequent entries of these cross elasticity vectors that flexibly incorporates the rates of decay across the previous entries.
\newline

\textbf{Definition 2}: \emph{Aggregate Competition}

One of the primary interests of the analysis is characterising products in terms of their sales sensitivities with respect to their competitors' prices. We introduce the notion of \it aggregate competition \rm (AC) to summarise the total effect of competition on a product's sales through its competitors' prices changes. We achieve this by defining the aggregate competition of product $i$ as the  sum of the top $l$ cross-elasticity coefficients. Concretely, the AC of a cross elasticity vector distribution is given by
$$ {\textrm{AC}}=\frac{1}{N}\sum_{i=1}^N\sum_{j=n-l_i+1}^{n} {{\eta}}_{i,\left(j\right)}.$$
The AC can be thought of as the total percentage effect that log price deviations of the top $l$ elasticity terms (where $l$ is the expected number of competitors terms) has with respect to the equivalent prices changes of the product's own log price. For example, if a product's AC is 0.25, it means  that if the log price decrease across each of its competitors was 1 unit, then the product's log price would need to decrease by 0.25 to offset the loss of sales its competitors prices changes would have had on the product's sales. Thus a large AC  indicates a product's sales are  significantly impacted by its competitors' prices.

\subsubsection{MCMC output}\label{subsubsec:MCMCoutputRetailData}
We present the MCMC output of the real retail cross-elasticity dataset and using the following priors  for $\left(\alpha,\beta,\lambda,w\right)$ and $ \nu$ (DP scale parameter): 
\begin{align*}
  \left(\alpha,\beta,\lambda,w\right) & \sim  Gamma\left(7, 7/10 \right) \times Gamma\left(0.5,1\right)\times Gamma\left(1,1 \right)  \times Beta\left(2,3\right) \\
  \nu & \sim Gamma\left(5,1\right)   
\end{align*}
respectively. These priors are selected to reflect a prior expectation of the decay and typical length of the cross-elasticity vectors in the retail analytics context. Specifically, the priors over $(\alpha,\beta) $ are selected to reflect prior knowledge of the modal nature of the coefficients and the expected heavy tailed nature of the cross elasticity coefficients. In addition, these priors were chosen more restrictive than in the simulated examples, to overcome the strong dependence between $\alpha$ and $\beta$ which, for small datasets such as this one, leads to weak identifiability. In particular, the prior for $\alpha$ is centred around $10$ as before, but with a variance of $\approx 14$. The complementary parameter $\beta$ is then centred at $0.5$ (corresponding to a heavy-tailed EW mixture), with a variance of $0.5$. In other words, mixtures are `shrunk' towards smaller values of $\beta$, i.e.,~towards assuming that there are no omitted competitors, unless the data strongly suggest otherwise.  The prior $w \sim Beta\left(2,3\right)$ is selected to prefer observed cross elasticities of length 4-5. The prior for $\lambda$ (purely a scale parameter) is uninformatively chosen as before, and the prior for $\nu$ is chosen such that, a priori, between 3 and 30 clusters (roughly) are expected. 

\begin{figure}[h]
  \centering
\includegraphics[ width=0.45\textwidth]{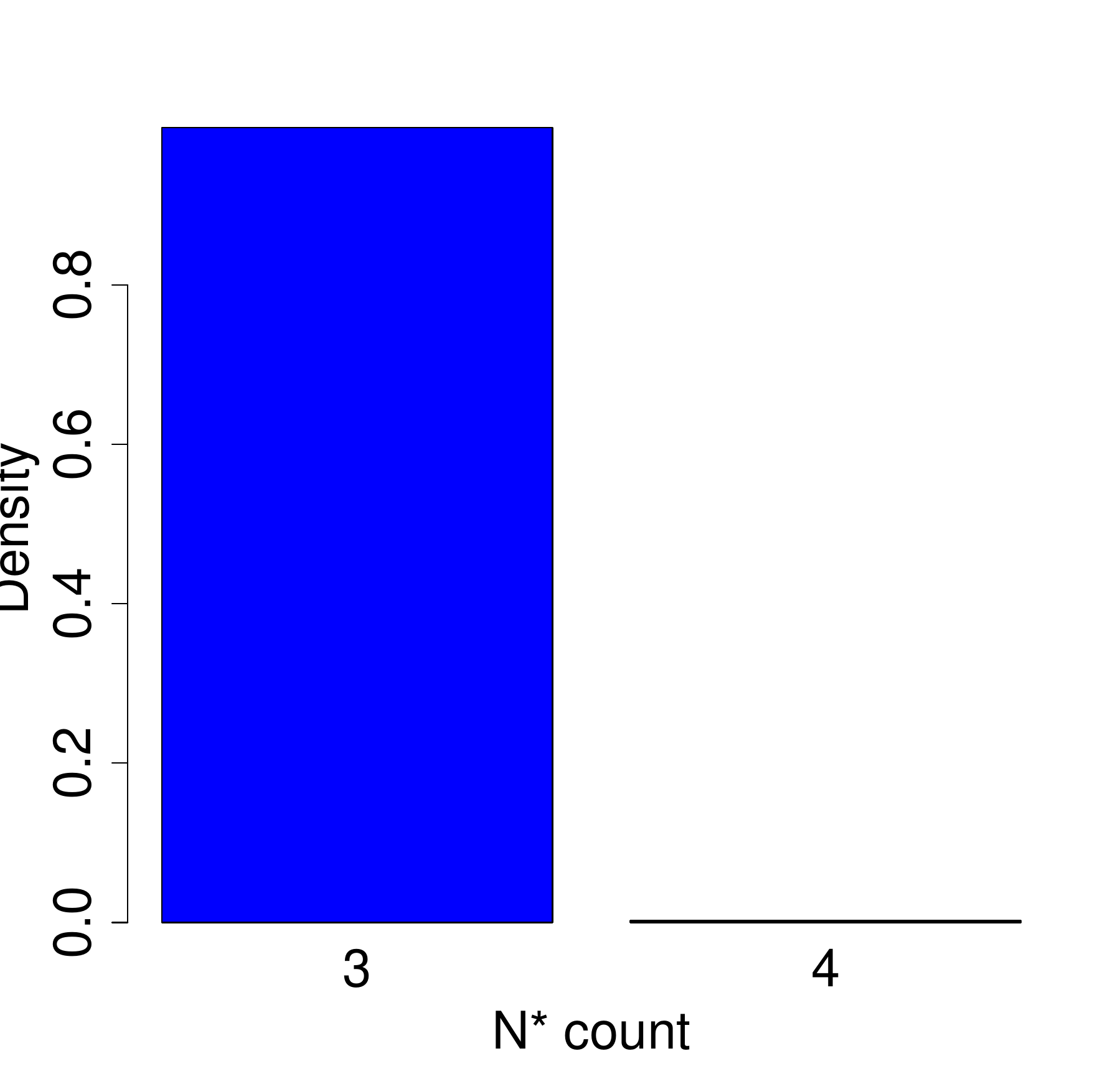}
\includegraphics[width=0.45\textwidth]{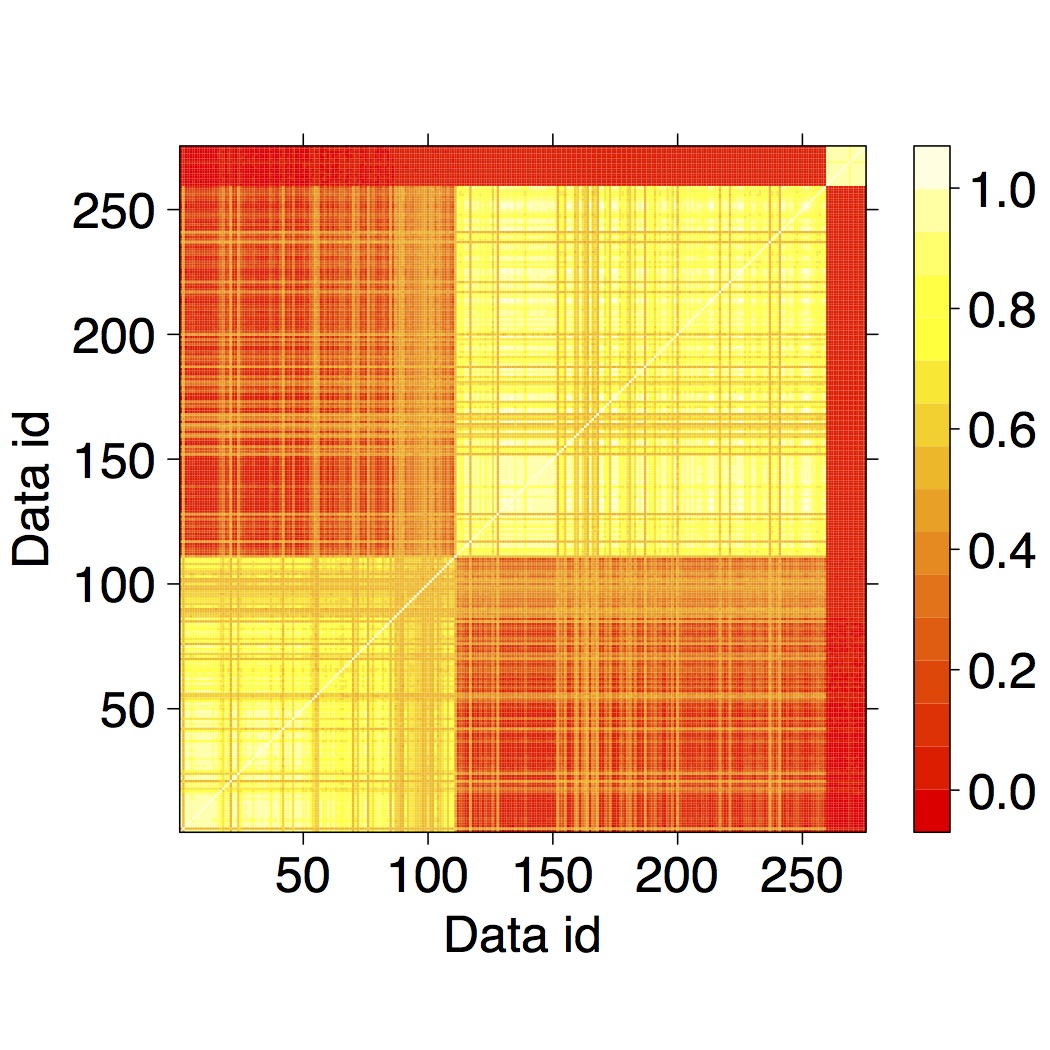}
 \caption{Left panel: histogram of $N^{*}$. Right panel: heatmap of cluster co-membership probabilities (re-grouped with respect to $\boldsymbol{C}^*$).}\label{plot:MCMCoutput_real}
\end{figure}

Figure \ref{plot:MCMCoutput_real} presents a histogram of the number of unique clusters $N^{*}$ and a heatmap of the posterior marginal coincidence probabilities (cluster co-membership). We see that the Maximum A Posteriori number of clusters is 3, with two large and one small cluster. Table \ref{tab:RA_summary} provides the category breakdown of each cluster, together with the number of observations in each as well as OC and AC values. It also includes the posterior mean and 2.5\% and 97.5\% posterior credible intervals of $\left(\alpha, \beta, \lambda, w \right)$ for each of  optimal clusters.

 Figure \ref{fig:RA_data} shows density estimates of the number of observed entries of the cross-elasticity vector, as well as the top two observed values in each vector, showing that our model is capturing these observed quantities very well. In particular, in Figure \ref{fig:RA_data} we observe a spike of very small values for the top order statistics $\eta_{(10)}$ and $\eta_{(9)}$, which the model is able to accommodate through a small value of $\alpha\beta$. 

\begin{figure}[h]
  \includegraphics[ width=0.31\textwidth]{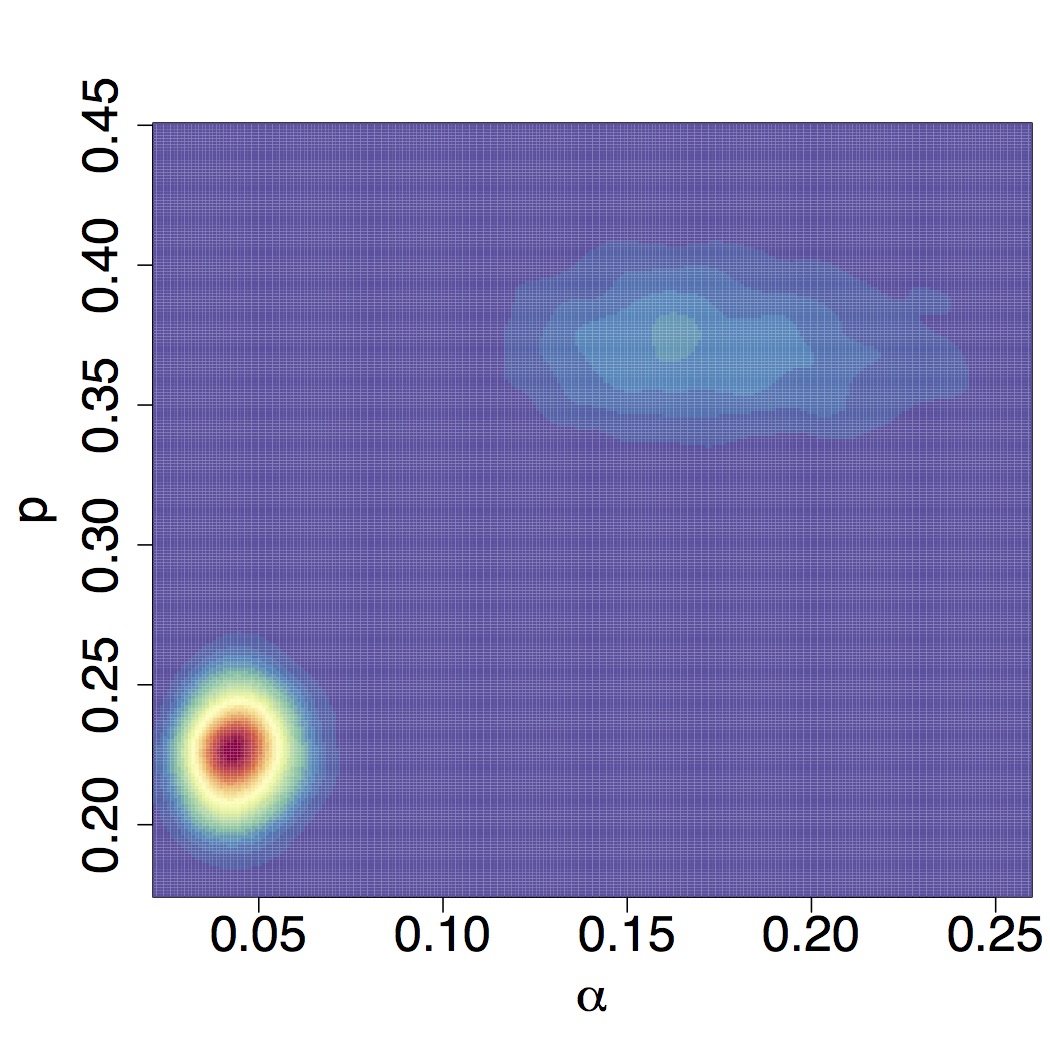}
  \includegraphics[ width=0.31\textwidth]{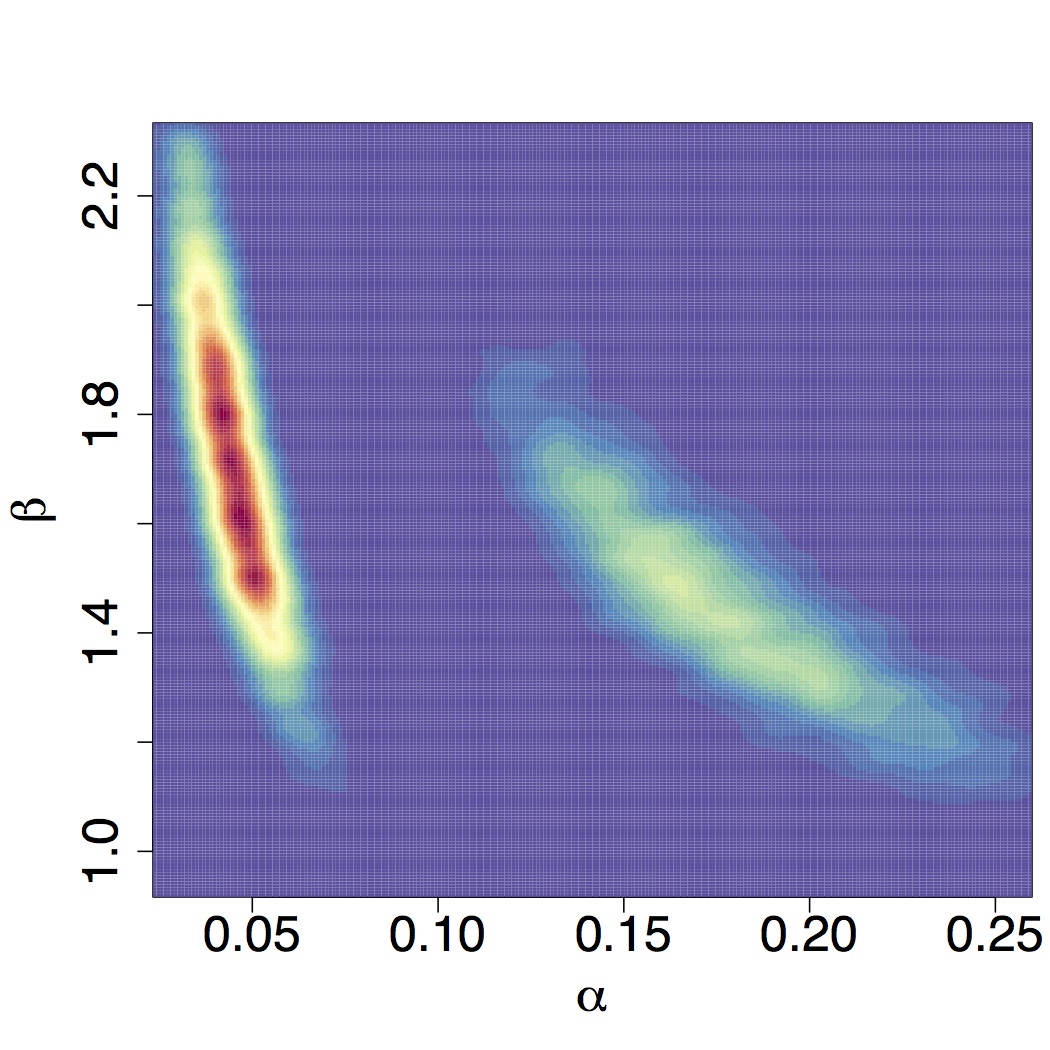}
  \includegraphics[ width=0.31\textwidth]{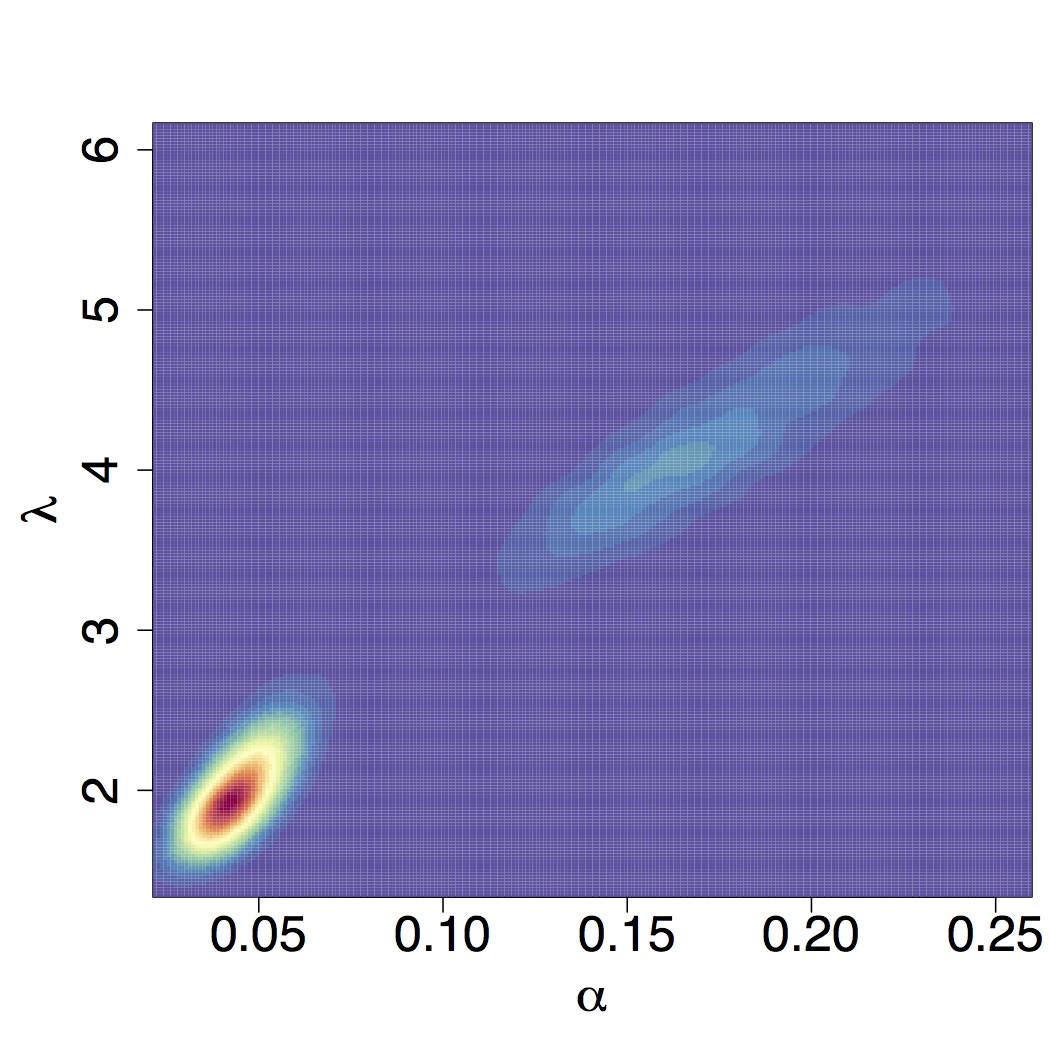}
  \\
  \includegraphics[ width=0.31\textwidth]{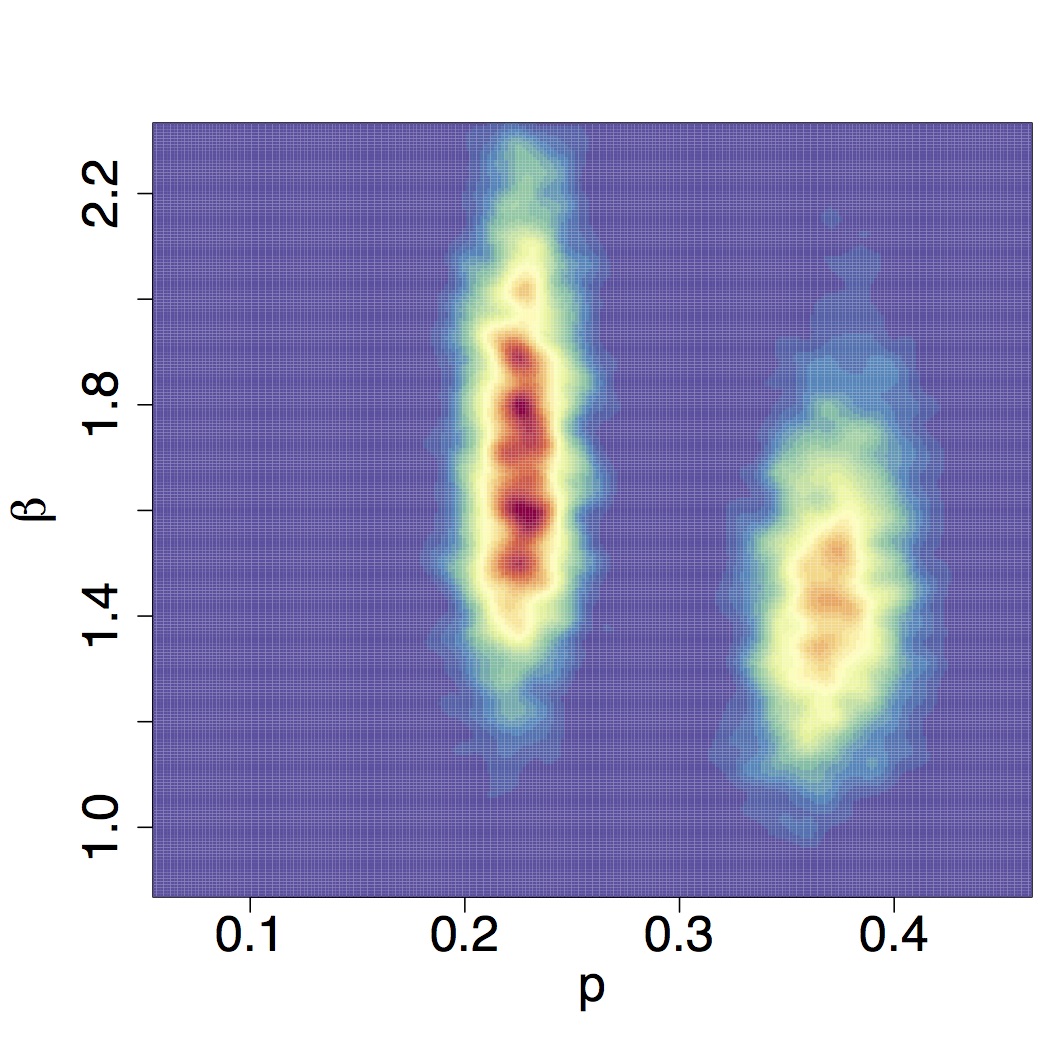}
  \includegraphics[ width=0.31\textwidth]{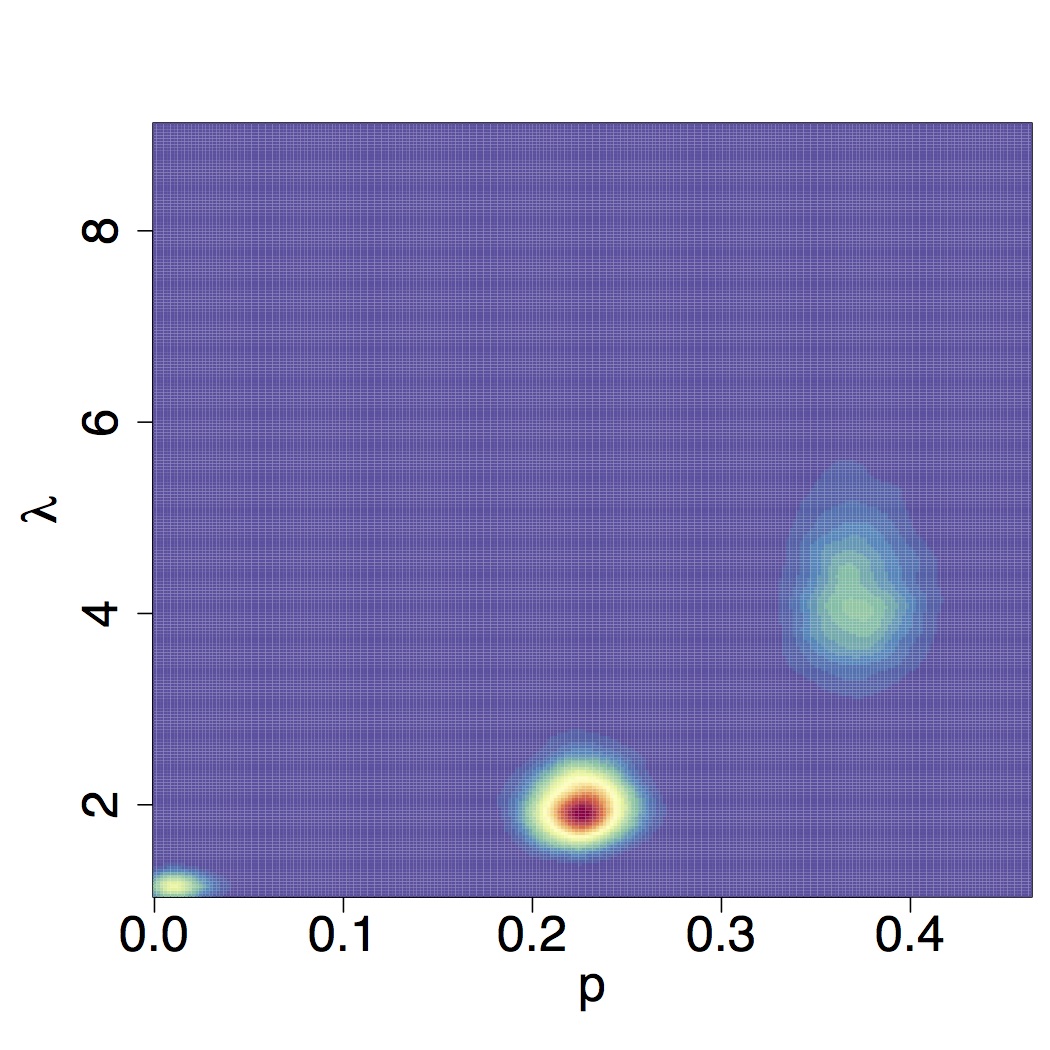}
  \includegraphics[ width=0.31\textwidth]{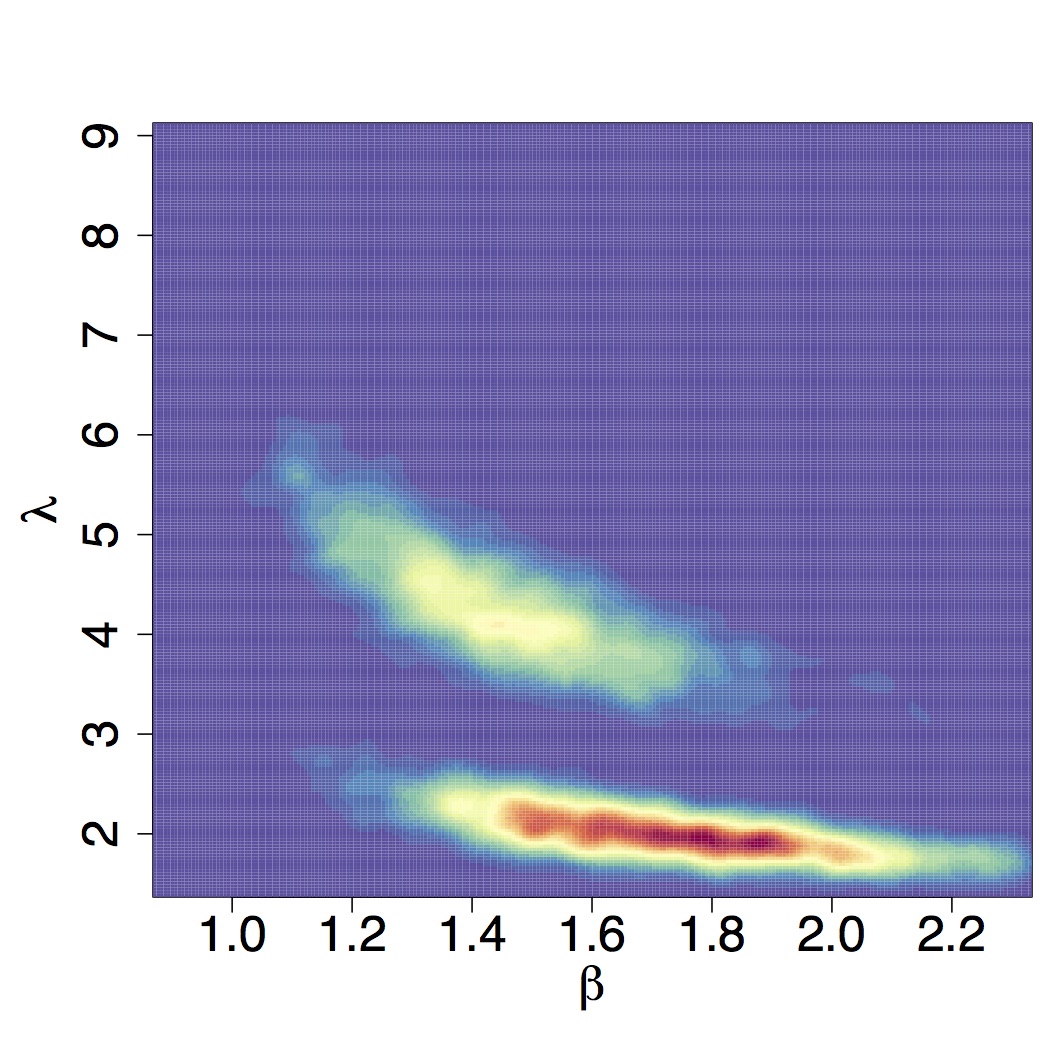}
  \caption{Heatmaps of pairwise posterior distributions of the parameters.}\label{plot:MCMCheatmaps_real}
\end{figure}

  Figure \ref{plot:MCMCheatmaps_real} provides heatmaps of pairwise posterior distributions of the parameters which demonstrate a neat separation between pairwise atoms. Interestingly, we observe that larger values of $\lambda$ (corresponding to a smaller mean) are associated with larger values of $w$; instead, $\beta$ values are inversely associated to values of $w$, suggesting that the censoring in this case is largely driven by $\beta$.

\begin{figure}[h]
  \centering
  \subfigure[ Cluster 1]{\includegraphics[ width=0.31\textwidth]{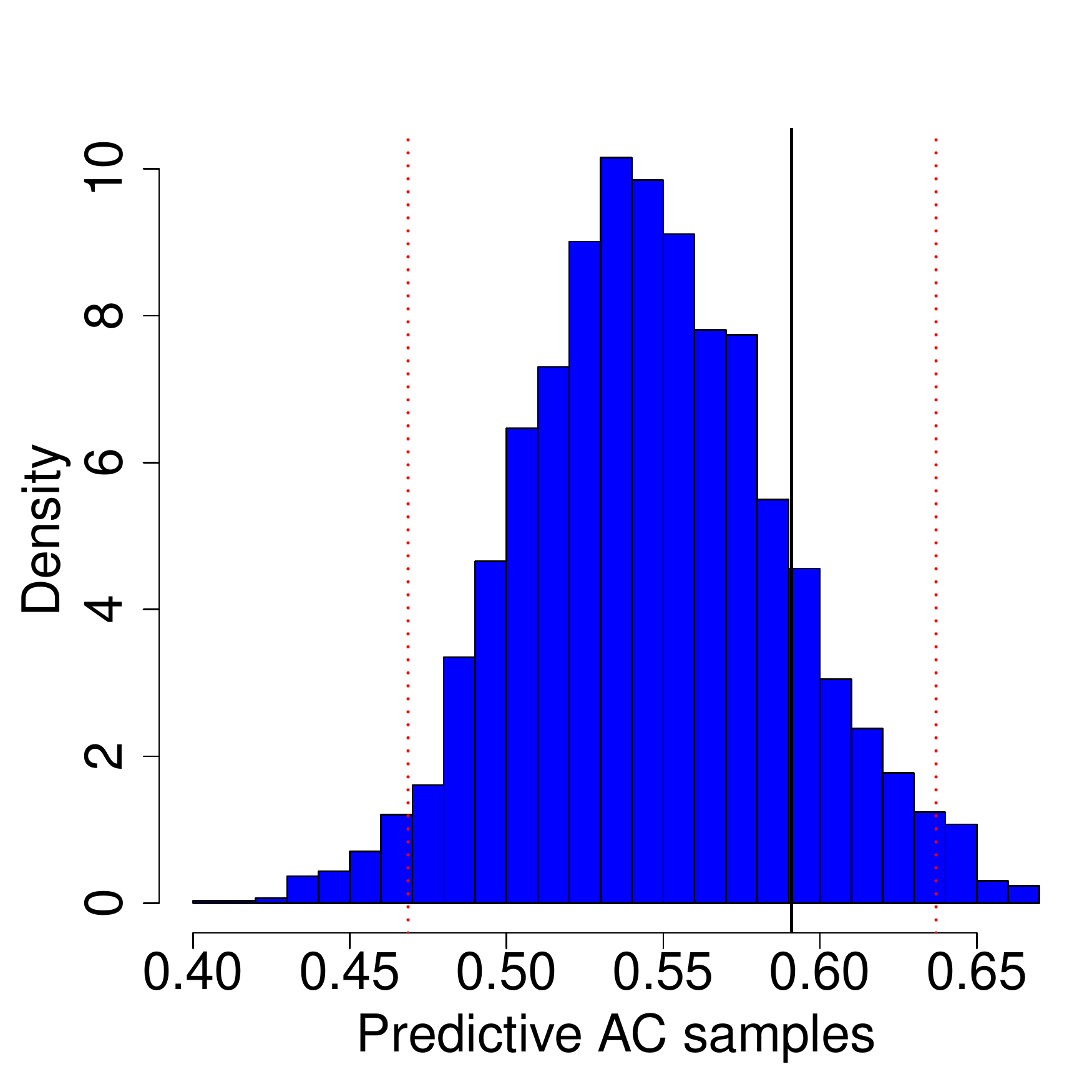}
  }
\subfigure[Cluster 2]{\includegraphics[ width=0.31\textwidth]{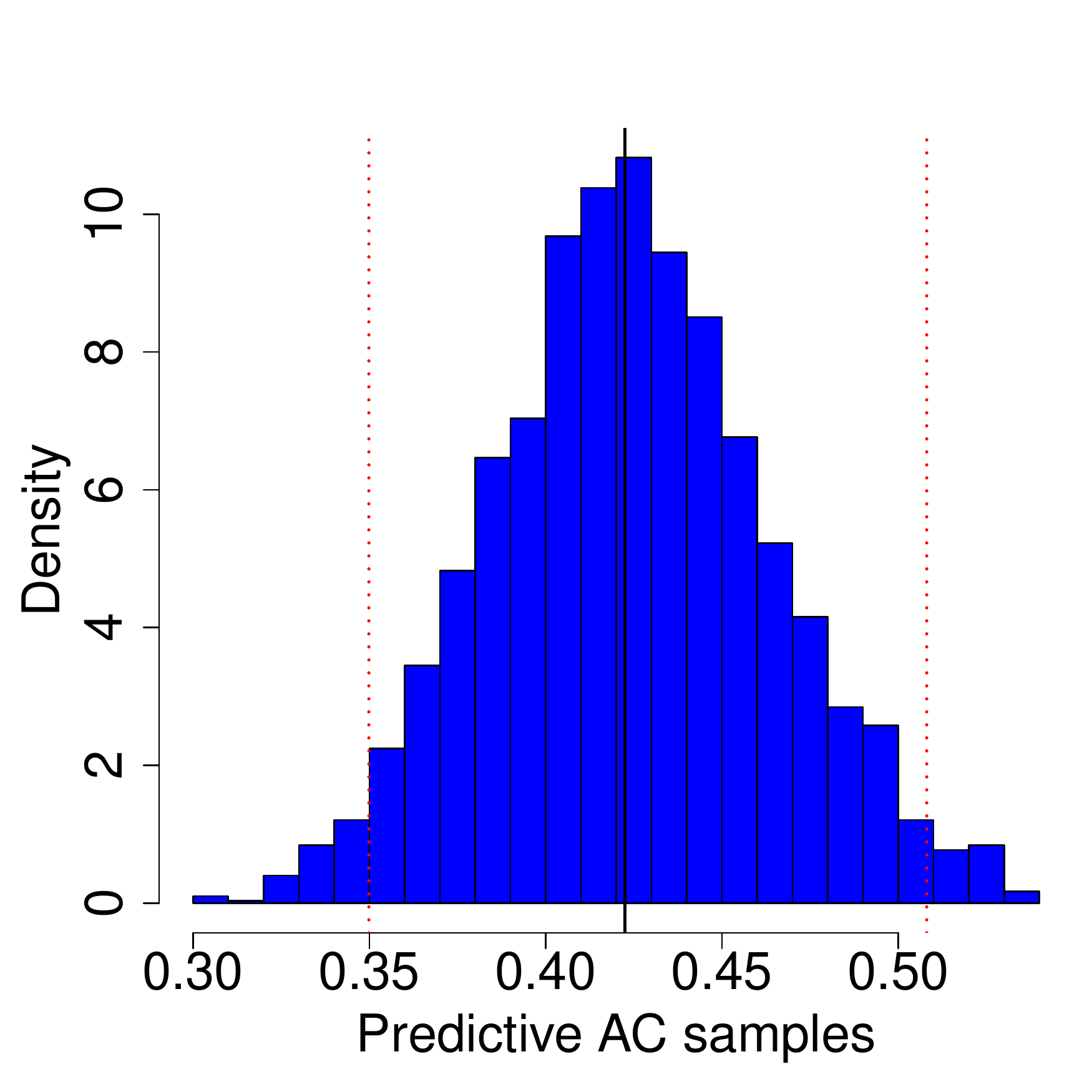}
  }
\subfigure[Cluster 3]{\includegraphics[ width=0.31\textwidth]{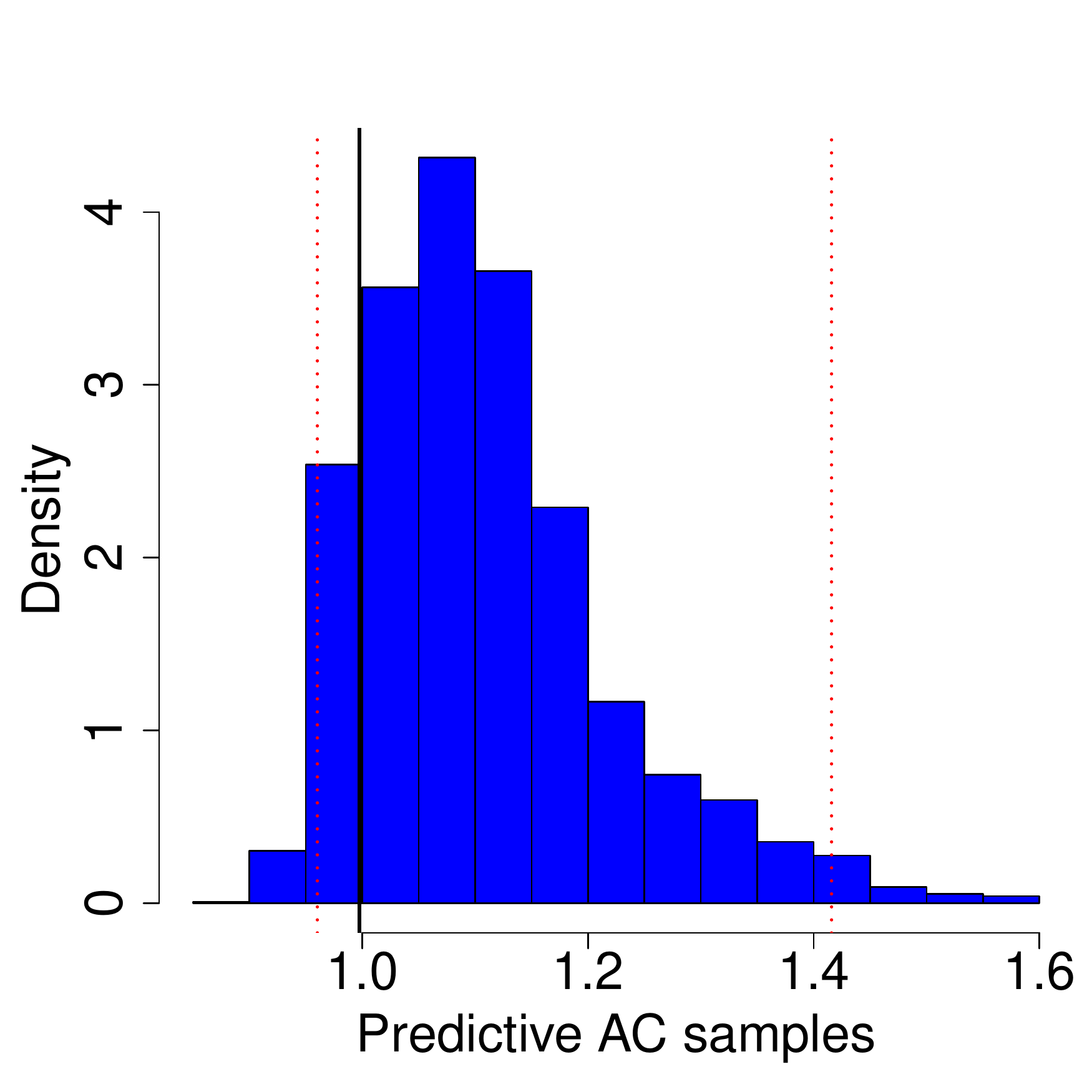}
  }
  \caption{\label{p_values} Histograms of  AC samples  with the 2.5\%, 97.5\% quantiles (red-dashed lines) and ${\textrm{AC}}$ (solid blue line) for each cluster in $\boldsymbol{C}^*$. We observe that our ${\textrm{AC}}$ test statistic  falls comfortably in the credible range for each cluster.}
\end{figure}

To assess model fit, we calculate the posterior predictive p-values \citep{Meng_1994} of $\textrm{AC}$ for each of the clusters defined by $\boldsymbol{C}^*$. Posterior predictive p-values  involves generating repetitions $ \textbf{\mbox{X}}^{rep}$ from the predictive distribution $ p\left(\textbf{\mbox{X}}^{rep} \mid \alpha,\beta, \lambda, w\right)$ for each MCMC sample  and calculating p-value $= 2\left(1-p\left(T\left(\textbf{\mbox{X}}^{rep}\right) >T\left(\textbf{\mbox{X}}\right) \mid \textbf{\mbox{X}}\right)\right) $ for some test statistic $T\left(\textbf{\mbox{X}}\right)$, in this case the aggregate competition. Figure \ref{p_values} provides predictive posterior p-values plots on the observed aggregate competition ${\textrm{AC}}$ over each cluster, compared against histograms of generated AC statistics over predictive replicates of $\textbf{\mbox{X}}$. These all comfortably fall within the 95\% prediction intervals. Trace plots of the $\left(\alpha,\beta, \lambda, w \right)$ atoms across the unique clusters are included in the Appendix \ref{sec:AppendixB}.


\begin{table}
  \caption{\label{tab:RA_summary}Retail analytics cluster-wise inference. Posterior means and (2.5\%,97.5\%) credible intervals for each of the four parameters $\left(\alpha, \beta, \lambda, w\right)$ along with other breakdown statistics for each the clusters.}
  \begin{tabular}{c c c c}
    \hline\hline
   Parameter & Cluster 1 & Cluster 2 & Cluster 3 \\ [0.5ex] 
    \hline\hline
    $\alpha$ & 0.16  (0.036,  0.28) &   0.06 (0.030, 0.21)    &    5.73  (2.49, 10.19)\\
    $\beta $ & 1.51 (1.10,  2.13) &      1.71 (1.20, 2.39)    &    10.88 (5.88 ,17.06) \\
    $\lambda$ & 3.89 (1.75, 5.81) &     2.29 (1.59, 4.47)   &  1.17 (1.09,   1.29)   \\
    $w $ & 0.34 (0.21, 0.41)          & 0.24 (0.20, 0.39)     & 0.016 (0.0017, 0.042)\\
    $N$ & 110 & 149 & 16 \\ 
    $ \textrm{OC} $ & 0.038 & 0.031  & 0.96 \\ 
    $\textrm{AC}$ & 0.55 & 0.42  &  1.11 \\ 
    
    \hline
    trad crisps  (22.5  \%)      & 30.9 \% & 18.1 \% & 6.25 \% \\ 
    exotic crisps    (33.1 \%)  & 33.6 \% & 35.6 \%  & 6.25 \% \\ 
    tortillas                		  (8.73  \%)     & 11.81 \%    & 7.38 \% & 0\% \\
    popcorn    	 (8.00 \%)  &    8.18 \%     &  8.05\%  &  6.25 \% \\
    nuts      		   (7.64 \%)  & 0\%  & 8.72 \%  & 50.0\%  \\ 
    dip         				 (4.73 \%)  & 4.55\%  & 5.37\%  & 0 \% \\ 
    pretzels   				 (2.18 \%)  &    0.909 \%    &  2.01 \% & 12.5\%  \\
    other       		   (13.1 \%)    &     10.00\%    &    14.8\%  & 18.8\% \\
    \hline

    $\mathbb{E}(\tilde{\eta}_{(10)})$ & 0.287 & 0.304 & 0.984 \\ 
    $\mathbb{E}(\tilde{\eta}_{(9)})$ & 0.156 & 0.124 & 0.963 \\ 
    $\mathbb{E}(\tilde{\eta}_{(8)})$ & 0.092 & 0.053 & 0.937 \\ 
    $\mathbb{E}(\tilde{\eta}_{(7)})$ & 0.055 & 0.022 & 0.926 \\ 
   
        \hline

  \end{tabular} 
  \end{table}

\subsubsection{\label{sec:ResultsCommentary}Retail analytics discussion }
Considering the clusters given by $\boldsymbol{C}^* $ and linking them to the corresponding categories, we see interesting breakdowns. Firstly, the first cluster has a high concentration of traditional flavoured crisps and no  nut products, whereas the second cluster has a significantly under average representation of traditional crisps. Finally the third cluster  comprises nuts, pretzels and the other product categories.

The first and second clusters appear not to have competitor products omitted from there regression models since $\textrm{OC}_{1}=0.038, \textrm{OC}_{2} =  0.031 <\epsilon$ and thus indicate that we do not expect any of the unobserved cross-elasticities to be of any significance. However, the third cluster exhibits competitor omission since $\textrm{OC}_{3}=  0.96 > \epsilon $. This implies that, according to the model, we expect to find at least one more competitor with a non-negligible cross-elasticity. 

The posterior mean values of parameters of the first cluster are  $\alpha_1=0.16,\,\beta_1=1.51$ with $w_1=0.34$ and an aggregate competition of $\textrm{AC}_{1}=0.55$, which point to a light-tailed distribution of cross-elasticities whose probability density diverges at 0. This is in line with the fact that this cluster largely consists of traditional crisps, which are a fiercely competitive product line,  where products have multiple substitutes and thus  a high degree of sales sensitivity is expected.  The second cluster exhibits similar behaviour, with posterior mean parameters $\alpha_2=0.06,\,\beta_2=1.71,w_2=0.24$ and an aggregate competition of $\textrm{AC}_{2}=0.42$, also implying a light-tailed distribution whose probability density diverges at 0. 

The third cluster is rather different; its posterior mean parameters $\alpha_3=5.73,\,\beta_3=10.88$ suggest a light-tailed distribution with mode away from 0. It largely consists of vectors with only a single cross-elasticity entry (through $w_3=0.016$), although the model suggests that an additional competitor may have been missed (or does not exist). Finally, its aggregate competition (despite the missing competitor) is $\textrm{AC}_{3}=  1.11$,  so that price changes of these leading competitor products can account for $1.11$ of equivalent prices changes of the product's own price changes. These parameters suggest that these products are substitutes, i.e. products  only bought as an alternative due to other equivalent products being unavailable or too expensive.

With respect to the expected values of the order statistic entries themselves, we observe similar order statistic patterns between the first and second clusters; each of the first order statistics entries accounts for a roughly similar amount of its leading direct elasticity ($28 \%$ and $30 \%$ respectively), however the decay rate between the subsequent order statistics of the first cluster is significantly slower than that of the second cluster (roughly $55 \%-60\%$ of their previous value compared with  $40 \%-45\%$). This decay rate observation between subsequent order statistics entries supports the discrepancy between each of the first and second cluster's AC statistics as well as the first cluster comprising of food items which traditionally have a high number of competitors than in the second cluster. Similarly as before, the third cluster differs significantly from the first and the second. Its first order statistic entry accounts for $98 \%$ of its leading direct elasticity and has a slower decay rate between successive order statistic sequences, each of these artefacts being significantly different from that of the previous clusters.

Retailers  also wish to understand the behaviour of their product range at a less granular level, e.g., at a category level. Clustering of cross-elasticity profiles provides a means to extract a new summary profile for a subset of products through a principled data-driven approach. 
Crucially, these can aid store planners and business specialists in the retail analytics domain to  better understand the optimal pricing and display combinations. For example, products in the third cluster are highly sensitive to specific competitor products, but otherwise are unaffected by the bulk of products around them. On the other hand, products in the first and second clusters are cannibalized by their competitor products, meaning that increasing the sale of one product decreases the sale of another, but with the second cluster being more robust to these prices changes than the first.

\section{\label{sec:Conclusion}Summary}
We have presented a Bayesian nonparametric mixture model for censored ordered data, using the Exponentiated Weibull distribution as a kernel. Our approach allows for flexible modelling of cross-elasticity coefficients without the need to specify the number of components and lends itself to meaningful interpretation. We implemented our methods on a dataset of cross-elasticities, focusing on quantities of interest in the retail analytics context, such as the aggregate competition and potential omitted competitors. Our model was able to capture several interesting features in the data through the corresponding clustering. 

These methods can potentially be extended in several directions. Firstly, one could introduce structure between the distribution of the length of the order statistics sequences and the kernel distribution. This may allow borrowing of information between these two sources of information, although it will become more computationally cumbersome. Secondly, one could relax the assumption of ordered observations to account for observations only ordered in expectation. Although in the cross-elasticity context this was not appropriate, in applications such as sports analytics it may be more reflective of the data. For example, the best athlete will not always have the best performance at a competition; instead, the ranking corresponds to average performance. Finally, we would like to explore combinations of different product categories to investigate similarities in market behaviour between otherwise disparate products.

\section*{Acknowledgements}
This work has been carried out with the financial support of the EPSRC, the Alan Turing Institute and dunnhumby ltd, our industrial partner. Access to the anonymised data servers was granted to us by  dunnhumby ltd. The anonymous dataset covers a wide range of sales categories and has sales going back over 5 years.

\appendix
\section{\label{sec:AppendixA}Posterior sampler}
Here we present the details of the posterior inference procedure. We provide more exact expositions of each of the three components of the MCMC algorithm; sampling $\boldsymbol{\theta}=\left(\alpha,\beta,\lambda,w\right)$ atoms of the $DP\left(\nu G_{0}\right)$ for each of the order statistics sequences, samples from $ p\left(\alpha,\beta,\lambda,w \mid \nu,\boldsymbol{x}_{\{i:\,C_i=k\}}\right)$ for each cluster and sampling the $\nu $.

\subsection{Sample from $ p\left({\theta}_i \mid {\theta}_{-i}, \nu, {x}_{i} \right)$}
As discussed during the posterior inference, we use the algorithm by \cite{neal2000markov}, we sample $\boldsymbol{\theta}_{i}=\left(\alpha_i,\beta_i,\lambda_i,w_i\right)$ by sampling from the multinomial distribution of degrees of freedom of order $N^{*}+c$ with entries
$$\boldsymbol{\theta}_{k}^{*} \overset{iid}{\sim} G_0 \text{ for } k=N^{*}+1,\ldots,N^{*}+c$$
$$ G_0 =  Beta\left(w \mid a, b\right) \times Gamma\left(\alpha \mid \alpha^1,\alpha^2\right) \times Gamma\left(\beta \mid \beta^1,\beta^2\right) \times Gamma\left(\lambda \mid \lambda^1,\lambda^2\right)$$
with probabilities $ P\left(\boldsymbol{\theta}_i = \boldsymbol{\theta}_{k}^{*} \mid \boldsymbol{\theta}_{-i}, \boldsymbol{x}_{i}, \boldsymbol{\theta}_{1}^{*},\ldots, \boldsymbol{\theta}_{N^{*}+c}^{*}\right)$, which is equivalent to
\begin{displaymath}
  P\left(C_i = k \mid C_{-i}, \boldsymbol{x}_{i}, \boldsymbol{\theta}_{1}^{*},\ldots, \boldsymbol{\theta}_{N^{*}+c}^{*}\right) \propto \left\{
    \begin{array}{lr}
      \frac{N^{*}_{k}}{N-1+\nu}f\left( \boldsymbol{x}_{i} \mid \boldsymbol{\theta}_{k}^{*} \right) \text{ for }1 \leq k \leq N^{*}\\
      \frac{\nu/k}{N-	1+\nu}f\left( \boldsymbol{x}_{i} \mid \boldsymbol{\theta}_{k}^{*} \right)  \  N^{*}< k \leq N^{*}+c
    \end{array}
  \right.
\end{displaymath}
 where
\begin{eqnarray}
f\left( \boldsymbol{x}_{i} \mid \boldsymbol{\theta}_{k}^{*} \right) &=& \binom{n-1}{l_i-1}\left(w_k^{*}\right)^{\left(l_i-1\right)}\left(1-w_k^{*}\right)^{\left(n-l_i\right)}\times \nonumber\\
&&\times   F\left(x_{i,\left(n-\left(l_i-1\right)\right)} \mid \alpha_k^{*}, \beta_k^{*}, \lambda_k^{*}\right)^{n-l_i} \prod_{j=1}^{l_i}f\left(x_{i,\left(n+j-l_i\right)} \mid \alpha_k^{*}, \beta_k^{*}, \lambda_k^{*}\right) 
\end{eqnarray}
where $ F\left(x \mid \alpha, \beta, \lambda\right) =  \left(1-e^{-\left(\lambda x\right)^{\beta}}\right)^{\alpha} $ and $ f\left(x \mid \alpha, \beta, \lambda\right) = \alpha \beta \lambda^{\beta} x^{\beta-1} \left(1-e^{-\left(\lambda x\right)^{\beta_i}}\right)^{\alpha-1} e^{-\left(\lambda x\right)^{\beta}} $. 

\subsection{Sample from $ p\left(\alpha^*,\beta^*,\lambda^*,w^* \mid \nu,{x}_{\{i:\,C_i=k\}}\right)$}
To ease notation, we suppress the asterisks from the exponents in this subsection.  For $t=1,\ldots,T$ iterations, for each unique cluster $k=1,\ldots,N^*$, we draw new parameters using an exponentiated Normal proposal for $\left(\alpha_k',\beta_k',\lambda_k'\right)$ centred at the points $\left(\log\left(\alpha_k^t\right),\log\left(\beta_k^t\right),\log\left(\lambda_k^t\right)\right)$ with standard deviations $\sigma_{\alpha}, \sigma_{\beta},\sigma_{\lambda}$ and a Normal proposal for $\left(w_k'\right)$ centred at the current point $\left(w_k'   \right)$ with standard deviation $\sigma_{w} $.
$$\left( \alpha_k', \beta_k', \lambda_k',w_k'\right)  \sim   \exp\left( N\left(\log\left(\alpha_k^t \right),\sigma^2_{\alpha}\right) \right) \times \exp\left( N\left(\log\left(\beta_k^t \right),\sigma^2_{\beta}\right) \right) \times  \exp\left(N\left(\log\left(\lambda_k^t \right),\sigma^2_{\lambda}\right) \right) \times N\left(w_k^t,\sigma^2_{w}\right)$$
This form of the proposals for $\left(\alpha_k',\beta_k',\lambda_k'\right)$ allows the scale of the proposal to vary according to the magnitude of the parameter values and ensures that proposed values are always positive. 
\newline
\newline
Then, if $0<w_k'<1$ (since $w$ is bounded between 0 and 1), with probability
$$ a =  min\left(1, \frac{\pi\left( \alpha_k', \beta_k', \lambda_k',w_k' \mid \boldsymbol{x}_{\{i:\,C_i=k\}}\right)}{\pi\left( \alpha_k^t, \beta_k^t, \lambda_k^t,w_k^t \mid \boldsymbol{x}_{\{i:\,C_i=k\}}\right)} \right),$$
 set $  \left( \alpha_k^{t+1}, \beta_k^{t+1}, \lambda_k^{t+1},w_k^{t+1}\right)=\left( \alpha_k', \beta_k', \lambda_k',w_k'\right)$, otherwise $ \left(\alpha_k^{t+1}, \beta_k^{t+1}, \lambda_k^{t+1},w_k^{t+1}\right)= \left(\alpha_k^t, \beta_k^t, \lambda_k^t,w_k^t\right)$. 
Here
\begin{eqnarray}
  \pi\left(\alpha_k,\beta_k,\lambda_k,w_k \mid \boldsymbol{x}_{\{i:\,C_i=k\}}\right) &\propto &  \alpha_k^{\alpha^{1}-1}e^{-\alpha^2 \alpha_k}\times  \beta_k^{\beta^{1}-1}e^{-\beta^2 \beta_k}  \times \lambda_k^{\lambda^{1}-1}e^{-\lambda^2 \lambda_k}\times w_k^{a-1}\left(1-w_k\right)^{b-1}\times \nonumber\\
&&\times \prod_{ \boldsymbol{x}_{i}:C_i=k} \Bigg[ w_k^{\left(l_i-1\right)}\left(1-w_k\right)^{\left(n-l_i\right)}  F\left(x_{i,\left(n-\left(l_i-1\right)\right)} \mid \alpha_k, \beta_k, \lambda_k\right)^{n-l_i}\times \nonumber\\
&&\phantom{ \prod_{ \boldsymbol{x}_{i}:C_i=k} \Bigg[}\times\prod_{j=1}^lf\left(x_{i,\left(n+j-l_i\right)} \mid \alpha_k, \beta_k, \lambda_k\right) \Bigg].\nonumber
\end{eqnarray}
The scales of the proposal normal distributions $\sigma_{w}, \sigma_{\alpha}, \sigma_{\beta},\sigma_{\lambda}$ should be tuned depending on the dataset.

\subsection{Sample from $ p\left(\nu \mid \alpha,\beta,\lambda,w,\textbf{\mbox{X}}\right)$}
Finally we implement the prior of  \cite{escobar1995bayesian} on $ \nu$ of the DP. By using the fact that by specify $ \nu \sim Gamma\left(\tau_1, \tau_2\right) $ and introducing an auxiliary variable $\gamma$ enables the sampling of $ \nu $ iteratively by a Gibbs sampler. Specifically, we take the following samples
$$ \left(\gamma \mid \nu, N^{*}\right) \sim Beta\left(\nu+1, N\right)$$
$$ \left(\nu  \mid \gamma, N^{*}\right) \sim \pi{_\gamma}Gamma\left(\tau_1+N^{*}, \tau_2-\log\left(\gamma\right)\right)+\left(1-\pi_{\gamma}\right)Gamma\left(\tau_1+N^{*}-1, \tau_2-\log\left(\gamma\right)\right)$$
where the weights $\pi_{\gamma}$ is defined by $\pi_{\gamma}/ \left(1-\pi_{\gamma}\right)=\left(\theta+N^{*}-1\right)/ \left(N\left(\tau_2-\log\left(\gamma\right)\right)\right)$. This concludes the a complete iteration of our posterior inference procedure.

 \section{\label{sec:AppendixB}Appendix. MCMC trace plots}
\begin{figure}[h]
   \centering
   \subfigure[$\sqrt\alpha$ trace posterior plots]
   {
   \includegraphics[width=0.3\textwidth]{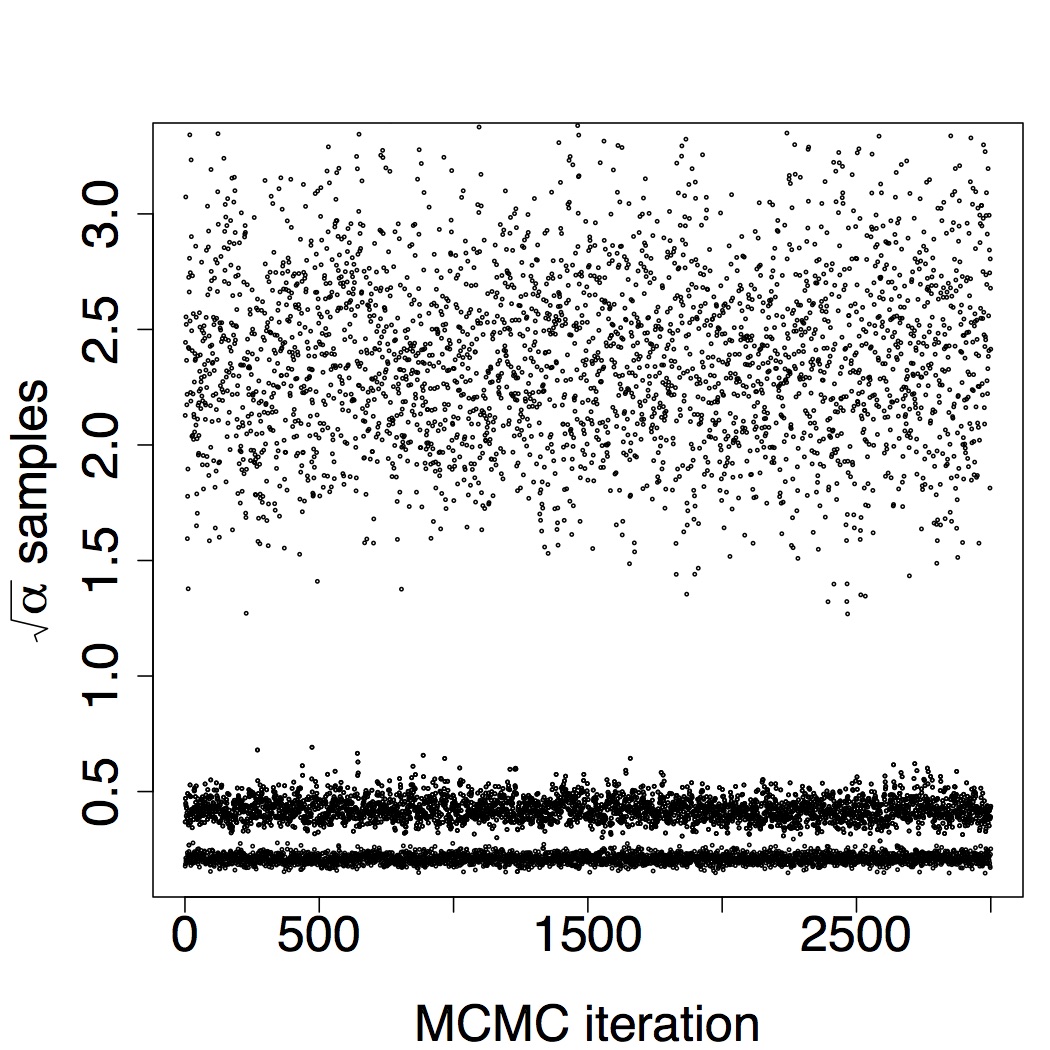}
   }
  \subfigure[$\sqrt\beta$ trace posterior plots]
  {
  \includegraphics[width=0.3\textwidth]{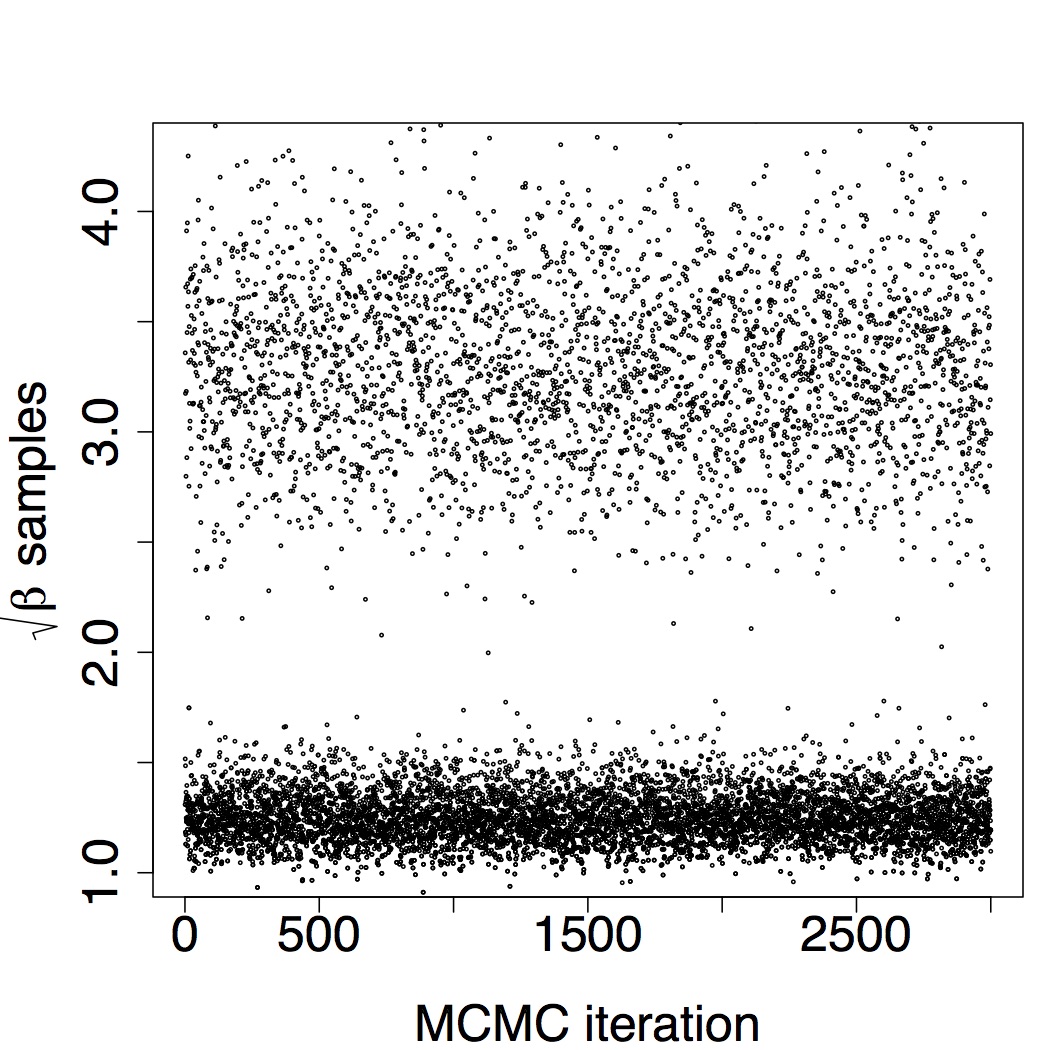}
  }
    \subfigure[$\sqrt\lambda$  trace posterior plots]
  {
  \includegraphics[width=0.3\textwidth]{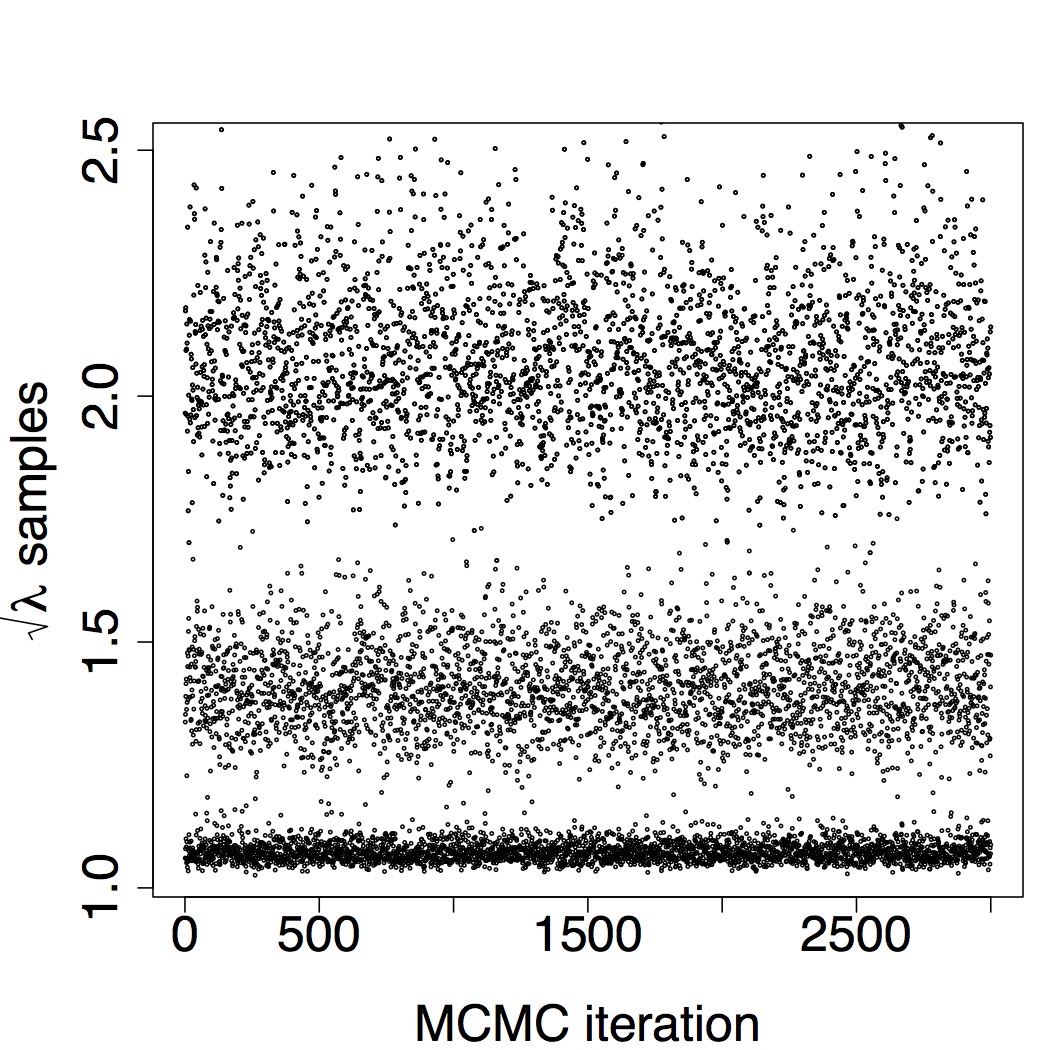}
  }
     \\
    \subfigure[$w$  trace posterior plots]
     {
       \includegraphics[width=0.3\textwidth]{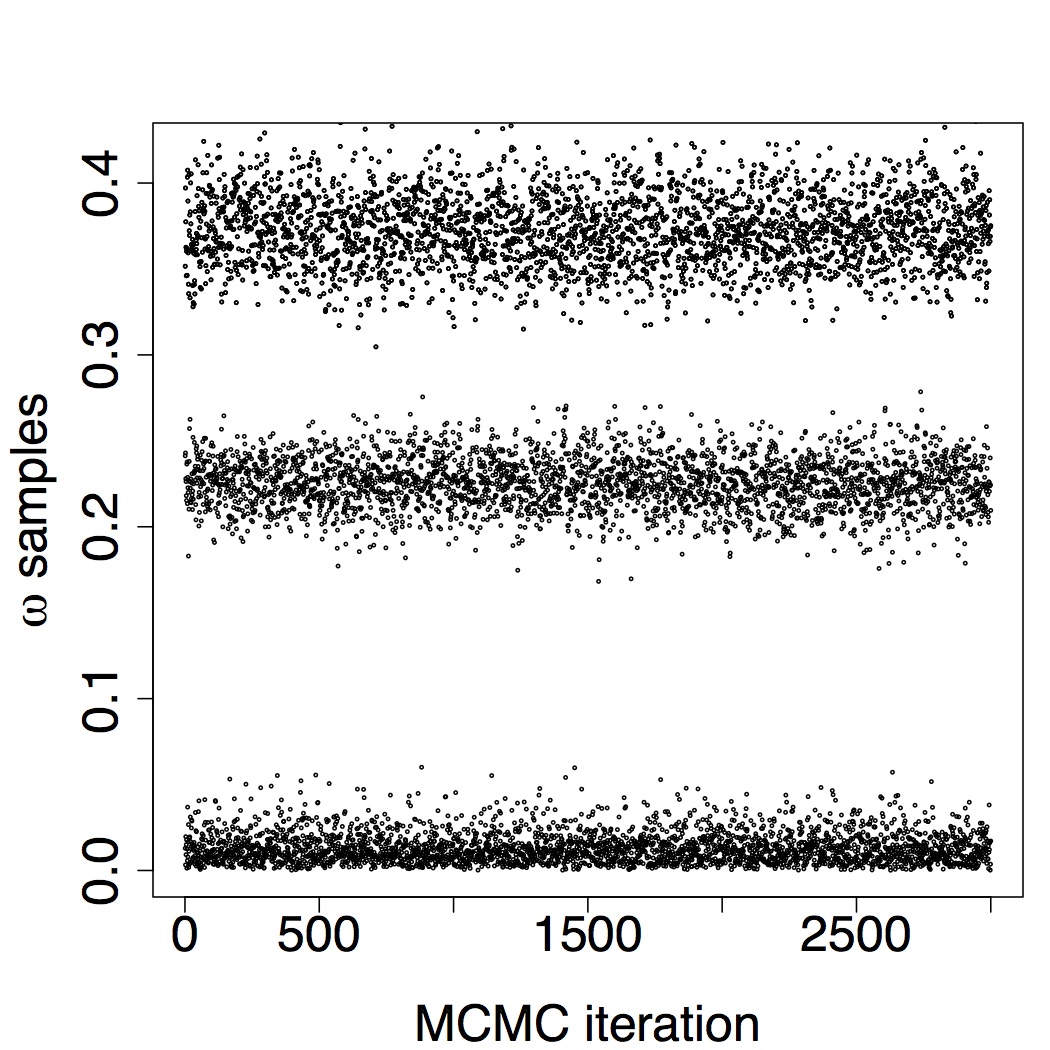}
    }
    \subfigure[$N^{*}$ trace posterior plots]
     {
       \includegraphics[width=0.3\textwidth]{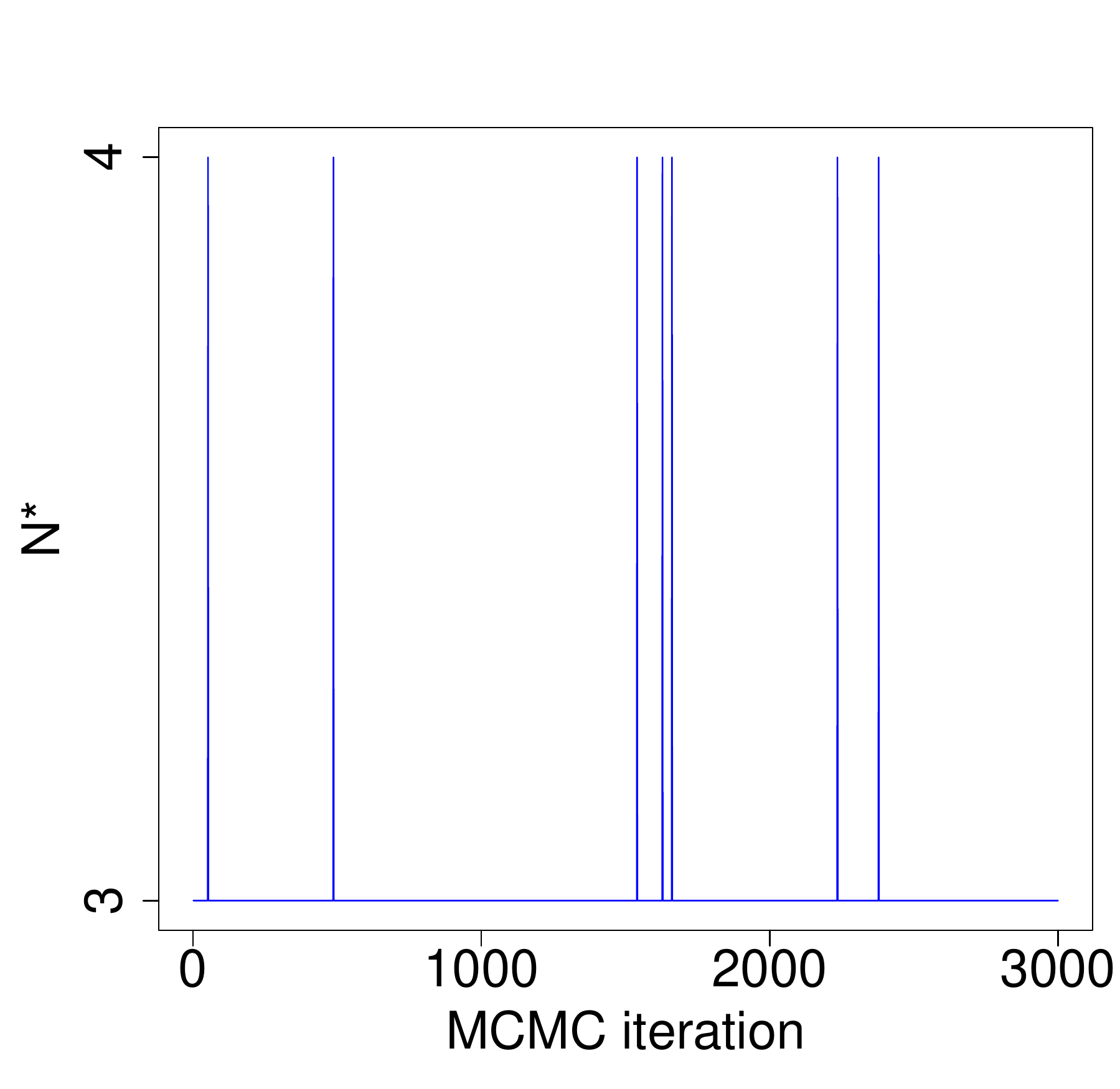}
    }
      \caption{Trace plots of MCMC samples for unique atoms of $\left(\sqrt\alpha,\sqrt\beta,\sqrt\lambda,w\right)$ parameters and $N^{*}$ on dunnhumby's cross elasticity data of the snack category.}\label{plot:MCMC_tracesl}
 \end{figure}

Here we assess the convergence of our inferential procedure on the retail analytics dataset.
Figure \ref{plot:MCMC_tracesl} provides traces of the atoms across all unique clusters  $\left(\left(\alpha^{*}_{k} \right)^{t}, \left(\beta^{*}_{k}\right)^{t}, \left(\lambda^{*}_{k}\right)^{t}, \left(w^{*}_{k}\right)^{t} \right)$ of $DP\left(\nu G_0 \right)$ samples for the iterations $t=1,\ldots,T$ across the unique atoms $k=1,\ldots, N^{*}_{t}$, where $N^{*}_{t}$ is the number of unique clusters at iteration $t$ and the trace of $N^{*}_{t}$. We plot the $\sqrt\cdot$ traces of $\left(\alpha, \beta, \lambda \right)$ to induce similar scales for graphical convenience. All plots indicate sufficient mixing and satisfactory convergence.

\bibliographystyle{plainnat}
\bibliography{Bibliography_thesis_JRSSC.bib}

\end{document}